\documentclass[a4paper,fleqn,usenatbib]{mnras}

\usepackage[T1]{fontenc}
\usepackage{ae,aecompl}
\usepackage{color}

\usepackage{graphicx}	
\usepackage{amsmath}	
\usepackage{amssymb}	
\usepackage{mathptmx}
\usepackage{txfonts}

\title[3rd July Outburst]{Evidence of sub-surface energy storage in comet 67P from the outburst of 2016 July 3}
\author[J. Agarwal et al.]{
\begin{minipage}{\textwidth}\Large
J. Agarwal,$^{1}$\thanks{E-mail: agarwal@mps.mpg.de (JA)}
V. Della Corte$^{2,3}$, 
P. D. Feldman$^{4}$, 
B. Gei\-ger$^{5}$, 
S. Merouane$^{1}$,
I. Bertini$^{6}$,
D. Bodewits$^{7}$,
S. Fornasier$^{8}$,
E. Gr\"un$^{9,10}$,
P. Hasselmann$^{8}$,
M. Hilchenbach$^{1}$,
S. H\"ofner$^{1}$,
S. Ivanovski$^{3}$,
L. Kolokolova$^{7}$,
M. Pajola$^{11}$,
A. Rotundi$^{2,3}$,
H. Sierks$^{1}$,
A. J. Steffl$^{12}$,
N. Thomas$^{13}$,
M. F. A'Hearn$^{7}$,
C. Barbieri$^{14}$,
M. A. Barucci$^{8}$,
J.-L. Bertaux$^{15}$,
S. Boudreault$^{1}$,
G. Cremonese$^{16}$,
V. Da Deppo$^{17}$,
B. Davidsson$^{18}$,
S. Debei$^{19}$,
M. De Cecco$^{11}$,
J. F. Deller$^{1}$,
L. M. Feaga$^{7}$,
H. Fischer$^{1}$,
M. Fulle$^{20}$,
A. Gicquel$^{18}$,
O. Groussin$^{21}$,
C. G\"uttler$^{1}$,
P. J. Guti\'errez$^{22}$,
M. Hofmann$^{1}$,
K. Hornung$^{23}$,
S. F. Hviid$^{24}$,
W.-H. Ip$^{25}$,
L. Jorda$^{20}$,
H. U. Keller$^{26}$,
J. Kissel$^{1}$,
J. Knollenberg$^{24}$,
A. Koch$^{27}$,
D. Koschny$^{27}$,
J.-R. Kramm$^{1}$,
E. K\"uhrt$^{28}$,
M. K\"uppers$^{5}$,
P. L. Lamy$^{29}$,
Y. Langevin$^{30}$,
L. M. Lara$^{22}$,
M. Lazzarin$^{14}$,
Z.-Y. Lin$^{25}$,
J. J. Lopez Moreno$^{24}$,
S. C. Lowry$^{31}$,
F. Marzari$^{14}$,
S. Mottola$^{25}$,
G. Naletto$^{6,17,32}$, 
N. Oklay$^{25}$,
J. Wm. Parker$^{12}$,
R. Rodrigo$^{33,34}$, 
J. Ryn{\"o}$,^{35}$,
X. Shi$^{1}$,
O. Stenzel$^{1}$,
C. Tubiana$^{1}$,
J.-B. Vincent$^{24}$,
H. A. Weaver$^{36}$,
B. Zaprudin$^{37}$
\end{minipage}
\\
\begin{minipage}{\textwidth}
\vspace{\baselineskip}
$^{1}$Max Planck Institute for Solar System Research, Justus-von-Liebig-Weg 3, 37077 G{\"o}ttingen, Germany;
$^{2}$Dipartimento di Scienze e Tecnologie, Universit\`a degli Studi di Napoli ``Parthenope'', CDN, IC4, 80143, Naples, Italy;
$^{3}$Institute for Space Astrophysics and Planetology, Via Fosso del Cavaliere 100, I-00133 Roma;
$^{4}$The Johns Hopkins University, Department of Physics and Astronomy, 3400 N. Charles Street, Baltimore, MD 21218, USA;
$^5$European Space Astronomy Centre, Camino bajo del Castillo s/n, 28692 Villanueva de la Ca\~{n}ada, Spain;
$^{6}$Centro di Ateneo di Studi ed Attivit\'a Spaziali "Giuseppe Colombo" (CISAS), University of Padova, Via Venezia 15, 35131 Padova, Italy;
$^{7}$Department for Astronomy, University of Maryland, College Park, MD 20742-2421, USA;
$^{8}$LESIA, Observatoire de Paris, PSL Research University, CNRS, Univ. Paris Diderot, Sorbonne Paris Cit\'e, UPMC Univ. Paris 06, Sorbonne Universit\'es, 5 Place J. Janssen, 92195 Meudon Pricipal Cedex, France;
$^{9}$Max Planck Institute for Nuclear Physics, Saupfercheckweg 1, 69117 Heidelberg, Germany;
$^{10}$Laboratory for Atmospheric and Space Physics, University of Colorado, 1234 Innovation Dr, Boulder, CO 80303, USA;
$^{11}$NASA Ames Research Center, Moffett Field, CA 94035, USA;
$^{12}$Southwest Research Institute, Department of Space Studies, Boulder, CO, USA;
$^{13}$Physikalisches Institut, Sidlerstrasse 5, University of Bern, CH-3012 Bern, Switzerland;
$^{14}$Department of Physics and Astronomy "G. Galilei", University of Padova, Vic. Osservatorio 3, 35122 Padova, Italy;
$^{15}$LATMOS, CNRS/UVSQ/IPSL, 11 Boulevard d'Alembert, 78280 Guyancourt, France;
$^{16}$INAF Osservatorio Astronomico di Padova, Vicolo dell'Osservatorio 5, 35122 Padova, Italy;
$^{17}$CNR-IFN UOS Padova LUXOR, Via Trasea 7, 35131 Padova, Italy;
$^{18}$Jet Propulsion Laboratory, M/S 183-401, 4800 Oak Grove Drive, Pasadena, CA 91109, USA;
$^{19}$Department of Industrial Engineering University of Padova Via Venezia, 1, 35131 Padova, Italy;
$^{20}$INAF - Osservatorio Astronomico di Trieste, via Tiepolo 11, 34143 Trieste, Italy;
$^{21}$Aix Marseille Universit\'e, CNRS, LAM (Laboratoire d'Astro-physique de Marseille) UMR 7326, 13388, Marseille, France;
$^{22}$Instituto de Astrofisica de Andalucia-CSIC, Glorieta de la Astronomia, 18008 Granada,  Spain;
$^{23}$Universit\"at der Bundeswehr LRT-7, Werner Heisenberg Weg 39, 85577 Neubiberg, Germany;
$^{24}$Institute of Planetary Research, DLR, Rutherfordstrasse 2, 12489 Berlin, Germany;
$^{25}$Institute of Astronomy, National Central University, 32054 Chung-Li, Taiwan;
$^{26}$Institute for Geophysics and Extraterrestrial Physics, TU Braunschweig, 38106 Braunschweig, Germany;
$^{27}$von Hoerner und Sulger GmbH, Schlossplatz 8, 68723 Schwetzingen, Germany;
$^{28}$Research and Scientific Support Department, European Space Agency, 2201 Noordwijk, The Netherlands;
$^{29}$Laboratoire d'Astrophysique de Marseille, UMR 7326 CNRS \& Aix-Marseille Universit\'e, 38 rue Fr\'{e}d\'{e}ric Joliot-Curie, 13388 Marseille cedex 13, France;
$^{30}$Instituto de Astrofisica de Andalucia-CSIC, Glorieta de la Astronomia, 18008 Granada,  Spain;
$^{31}$Centre for Astrophysics and Planetary Science, School of Physical Sciences, The University of Kent, Canterbury CT2 7NH, United Kingdom;
$^{32}$Department of Information Engineering, University of Padova, Via Gradenigo 6/B, 35131 Padova, Italy;
$^{33}$Centro de Astrobiologia (INTA-CSIC), European Space Agency, European Space Astronomy Centre (ESAC), P.O. Box 78, E-28691 Villanueva de la Canada, Madrid, Spain;
$^{34}$International Space Science Institute, Hallerstrasse 6, 3012 Bern, Switzerland;
$^{35}$Finnish Meteorological Institute, Climate Research, Erik Palmenin aukio 1, P.O. Box 503, FI-00101 Helsinki, Finland;
$^{36}$Johns Hopkins University Applied Physics Laboratory, Laurel, MD 20723, USA;
$^{37}$University of Turku, Department of Physics and Astronomy, Tuorla Observatory, V\"ais\"al\"antie 20, 21500 Piikki\"o, Finland
\end{minipage}
}
\date{Accepted XXX. Received YYY; in original form ZZZ}
\pubyear{2017}
\begin{document}
\label{firstpage}
\pagerange{\pageref{firstpage}--\pageref{lastpage}}
\maketitle

\begin{abstract}
On 3 July 2016, several instruments on board ESA's Rosetta spacecraft detected signs of an outburst event on comet 67P, at a heliocentric distance of 3.32\,AU from the sun, outbound from perihelion. We here report on the inferred properties of the ejected dust and the surface change at the site of the outburst. The activity coincided with the local sunrise and continued over a time interval of 14 -- 68 minutes. It left a 10m-sized icy patch on the surface. The ejected material comprised refractory grains of several hundred microns in size, and sub-micron-sized water ice grains.
The high dust mass production rate is incompatible with the free sublimation of crystalline water ice under solar illumination as the only acceleration process. Additional energy stored near the surface must have increased the gas density. We suggest a pressurized sub-surface gas reservoir, or the crystallization of amorphous water ice as possible causes.
\end{abstract}




\section{Introduction}
Outbursts are sudden and short-lived events of mass loss from the surfaces of comets. They have been observed in many comets, on different scales, and under various circumstances. A variety of models have been developed to explain their appearance \citep{hughes1990,belton2010}. Repeating early-morning outbursts at specific sites on comet 9P/Tempel 1 have been attributed to the re-sublimation of water ice frozen out in the uppermost surface layer during the preceding night, as sublimation in deeper layers would continue during night due to the delay between sub-surface and surface temperature cycles \citep{prialnik-ahearn2008}. Outbursts uncorrelated with local time can be driven by cryo-volcanism, following the crystallization of amorphous water ice in the deep ($\sim$15\,m) interior and the release of trapped CO or CO$_2$ \citep{belton-feldman2008}. Also the deepening of a pre-existing crack into layers containing highly volatile material has been proposed as the cause of some outbursts \citep{skorov-rezac2016}. 
Collapsing sub-surface voids formed by the earlier sublimation of a volatile substance \citep{vincent-bodewits2015} and collapsing cliffs create dust clouds that can be perceived as outbursts \citep{steckloff-graves2016,vincent-ahearn2016,pajola-hoefner2017}. 

The 2.5-year Rosetta mission at comet 67P/Churyumov-Gerasimenko witnessed a large number of outbursts on various scales. In a catalogue of all optically detected outbursts during the perihelion passage, \citet{vincent-ahearn2016} found that the events cluster into two groups by the local time of their appearance: one group occured in the early morning and was attributed to the rapid change in temperature and resulting thermal stress, the other group was observed in the early afternoon and attributed to the diurnal heat wave reaching a deeper layer enriched in volatiles.

Most outbursts from comet 67P were detected only by a single instrument \citep{knollenberg-lin2016, feldman-ahearn2016, vincent-ahearn2016}. For many, the approximate source region on the ground could be reconstructed \citep{vincent-ahearn2016}, but no systematic search for the traces of the induced surface change has been performed yet, and it might prove difficult in many cases due to the uncertainty of the source region coordinates. 

In a few events, Rosetta coindicentally flew through the plume of ejected material, while the outburst was also, serendipitously, documented by one or several remote sensing instruments. These events provide particularly valuable data sets due to the nearly simultaneous measurements of several instruments putting strong constraints on the properties of the ejected material and the temporal evolution of the outburst process. Such multi-instrument observations of an outburst on 2016, February 19 were analysed in \citet{gruen-agarwal2016}. Unfortunately, the location of the site of origin of that outburst could not be derived with certainty. On the other hand, \citet{pajola-hoefner2017} could study in great detail the surface change induced by the collapse of a cliff, while only little data on the ejected material were available.

The topic of the present paper is an outburst that occured on 2016, July 03, and was detected by at least 5 instruments on board Rosetta, such that the quantity, composition, and velocities of the ejected material can be derived with some certainty. In addition, the point of origin of this event, its topographic conditions, and the induced surface change can be studied in detail due to the serendipitous availability of high-quality images. 

In the following Section~\ref{sec:data}, we briefly decribe the location and timing of the event, followed by detailed accounts of the measurements of the individual instruments and their interpretation. In Section~\ref{sec:results}, we derive properties of the ejected material and discuss the topography and topographic change at the outburst site. In Section~\ref{sec:processes}, we discuss possible processes to trigger the outburst, and in Section~\ref{sec:summary} we summarize the key findings and discuss their significance for a larger context.

\section{Measurements and their interpretation by instrument}
\label{sec:data}

\begin{figure}
\centering
\includegraphics[width=0.48\columnwidth]{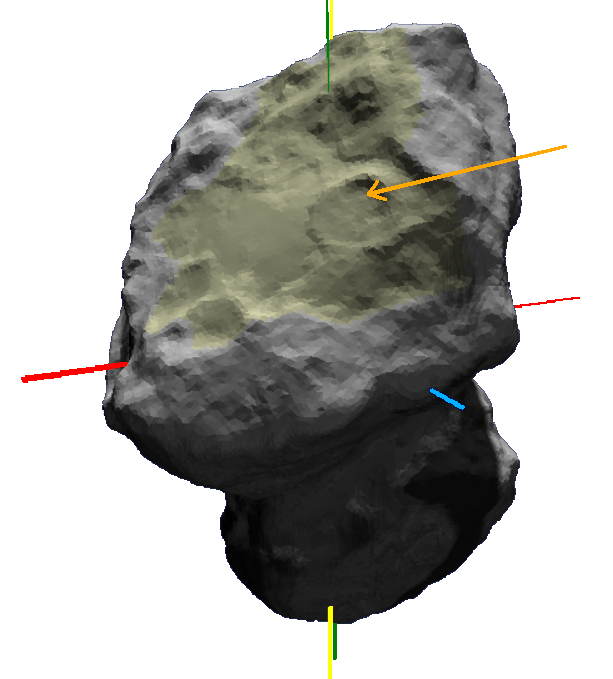}
\includegraphics[width=0.48\columnwidth]{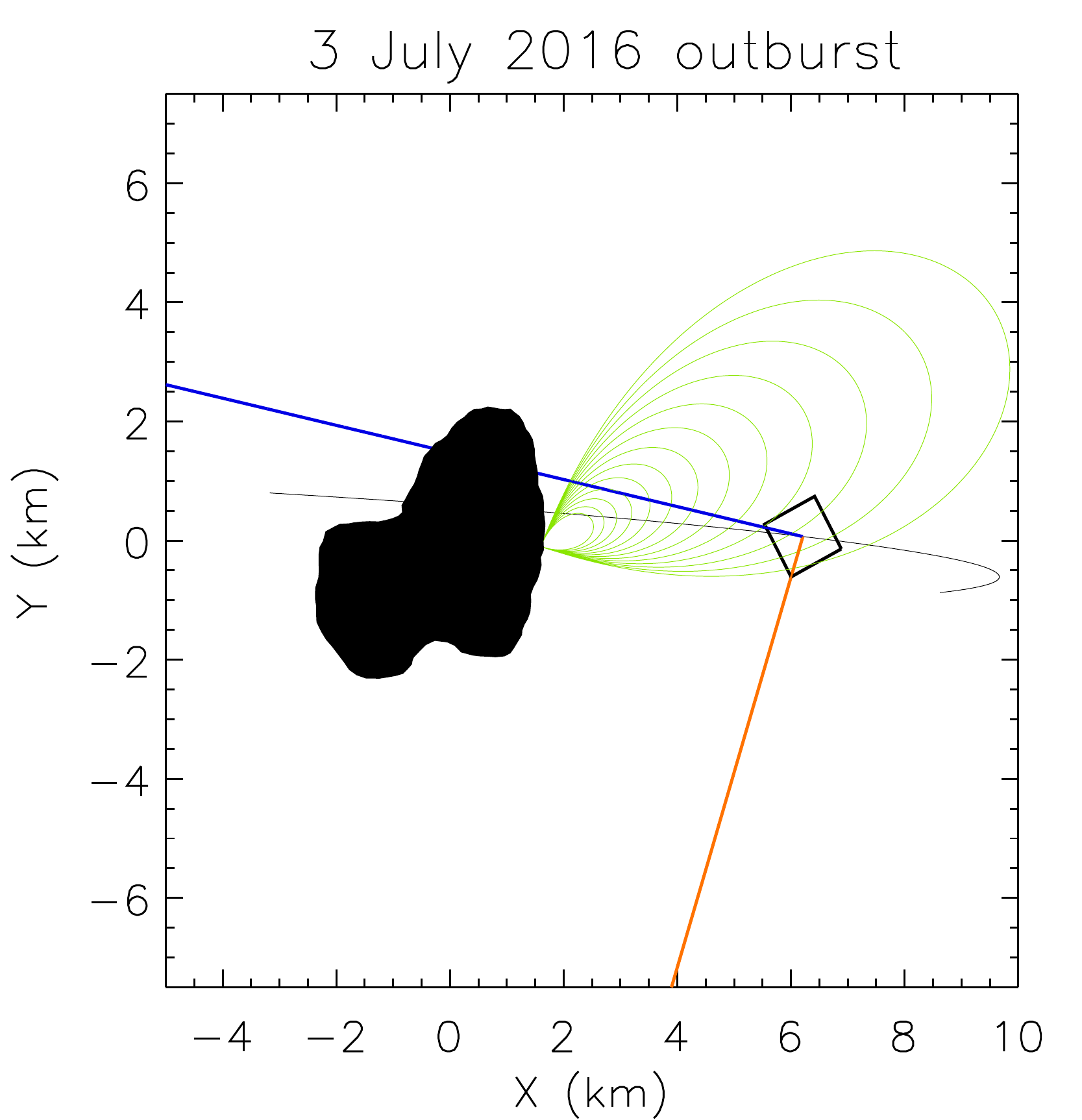}
\caption{Left panel: Location of the outburst site on the comet. The model shows the comet as seen from Rosetta on 2016 July 03 at  07:50. The arrow marks the outburst site in the circular Basin F in the Imhotep region (shaded yellow). The yellow lines indicate the comet-Sun line with the Sun at the top of the image. The green, red, and light blue lines mark the x-, y-, and z-axis of the comet. The south pole is visible where the z-axis crosses the comet surface. The image was generated with the CG viewer tool developed by Tim Wittrock (https://planetgate.mps.mpg.de:8114). Right panel: approximate configuration of the plume and the spacecraft. The blue line represents the boresights of the science instruments, the orange line that of STR-B.} 
\label{fig:geometry}
\end{figure}

On 2016, July 3, comet 67P was at a distance of 3.32 astronomical units (AU) from the Sun, outbound from its perihelion passage on 2015, August 13. Rosetta was in a close orbit about the nucleus, at a distance of 8.5 km from the outburst site that is located inside the circular ($\sim$500\,m diameter) Basin F \citep{auger-groussin2015} in the Imhotep region in the southern hemisphere of comet 67P, at 172.0$^\circ$ longitude and -33.2$^\circ$ latitude (Fig.~\ref{fig:geometry}). The Rosetta instruments contributing to this work and their data concerning the outburst are summarized in \ref{tab:instruments}

\begin{table}
\centering
\label{tab:instruments}
\caption{List of instruments contributing to this work. The first and second columns list the names of the instruments and the type of measurement they perform. The third column lists the type of data contributed to this work.}
\begin{tabular}{lll}
\hline
Name & Type & Data \\
\hline
Alice & Far-UV imaging spectrograph & Spectrum of the plume\\
COSIMA & TOF mass spectrometer & Microscope images of \\
&&1 fragmented particle\\
GIADA & Dust detector & Mass and velocity of \\
&& 22 particles\\
OSIRIS & Camera system & (Colour-)Images of the \\
&&outburst site \\
STR-B & Star tracker camera & Brightness of the \\
&& diffuse dust background\\
\hline
\end{tabular}
\end{table}

\begin{figure*}
\includegraphics[width=\textwidth]{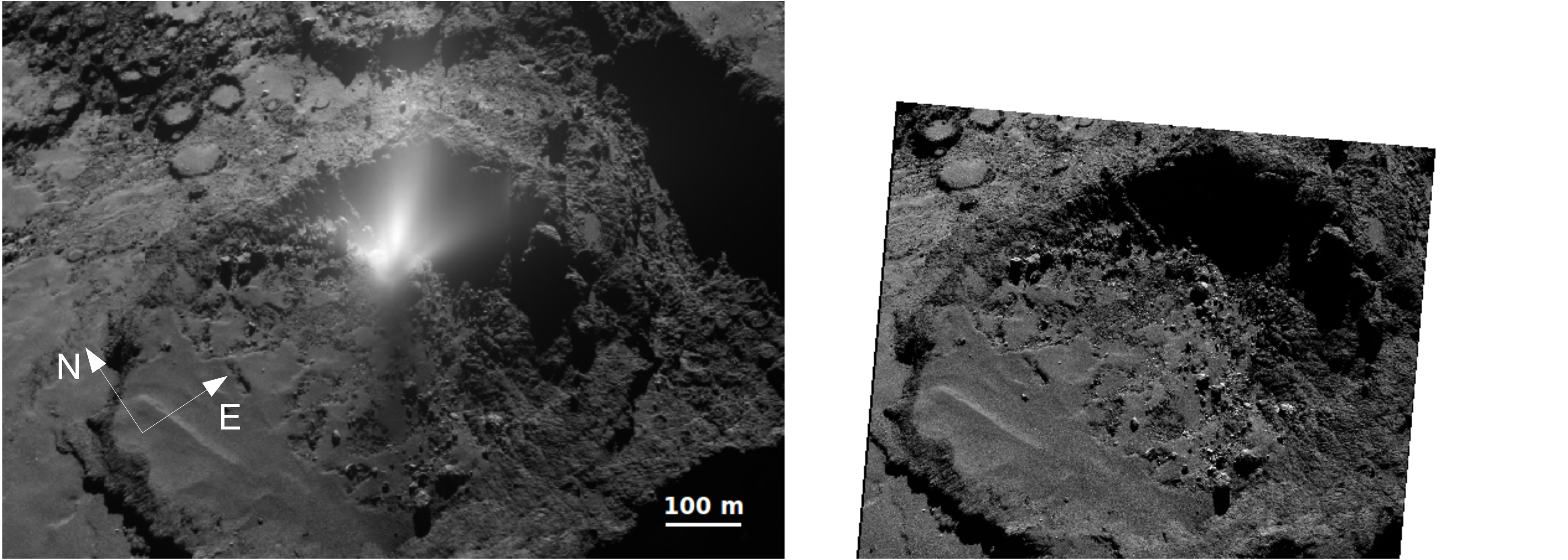}\\[0.2\baselineskip]
\includegraphics[width=\textwidth]{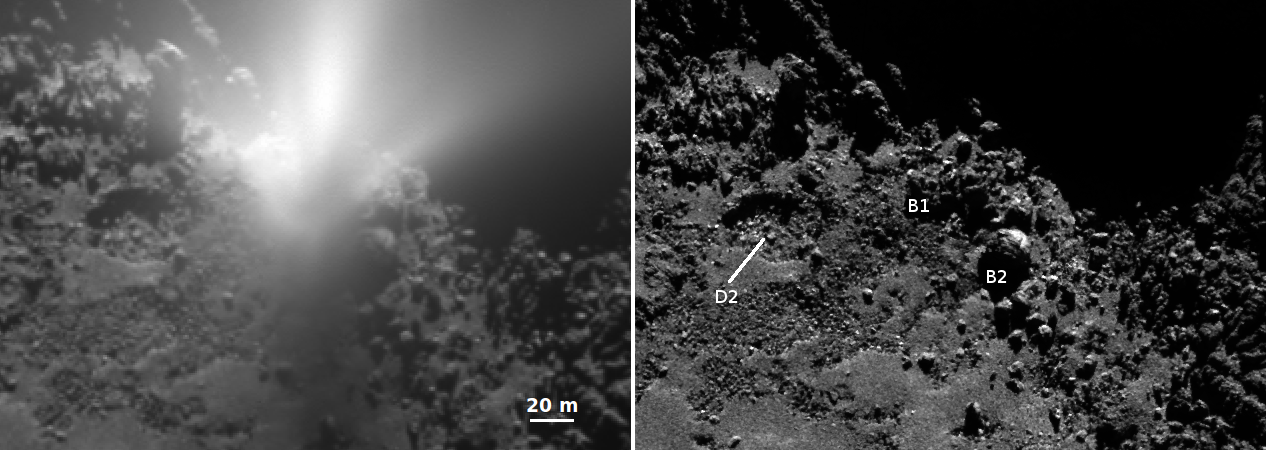}
\caption{Left: WAC image of the outburst plume obtained on  2016 July 03 7:50. Right: the same region at the same scale and under similar viewing conditions observed with NAC on May 03. The upper and lower row show the same images at different zoom levels. The boulders B1 and B2, and a neighbouring depression D2 are indicated for orientation and comparison to Fig.~\ref{fig:mar19-jul02}.
}
\label{fig:outburst_site}
\end{figure*}

A dust plume and its point of origin on the surface were observed by the Ultraviolet Imaging Spectrograph, Alice, beginning from 07:36 and by the Wide Angle Camera (WAC) of the Optical, Spectroscopic, and Infrared Remote Imaging System (OSIRIS) at UT 07:50 (Fig.~\ref{fig:outburst_site}). Unless specified otherwise, all further times refer to UT on 2016, July 3.
A WAC image of the same region from 07:04 shows the site still in shadow and no sign of dust activity near it. The background radiance of the star tracker camera STR-B began to increase around 07:40, and the Grain Impact Analyzer and Dust Accumulator (GIADA) detected the first particle at 08:26. The peak flux in STR-B and GIADA was observed between 08:40 and 09:00. Material from the outburst was also detected on the dust accumulation targets of the COmetary Secondary Ion Mass Analyzer (COSIMA).
An OSIRIS image of the outburst site obtained at 08:48 does not show any obvious dust near the outburst site. GIADA detected the last particle at 10:29, while the background signal of STR-B had not reached its pre-outburst level at 14:00 and continued to decline.  

\subsection{OSIRIS}
\label{subsec:osiris}
OSIRIS \citep{keller-barbieri2007} onboard the Rosetta spacecraft comprised a Narrow- and a Wide Angle Camera (NAC and WAC), each with a CCD detector of 2048 $\times$ 2048 pixels. The fields of view (FOVs) covered approximately 2$^\circ \times$ 2$^\circ$ and 12$^\circ \times$ 12$^\circ$, respectively. The cameras were regularly imaging the nucleus and coma of comet 67P/Churyumov-Gerasimenko between 2014, March and 2016, September in 25 broad- and narrow-band filters covering the wavelength range 240 to 1000 nm \citep{sierks-barbieri2015}. The standard data processing on ground comprised bias-subtraction, flat-fielding, correction for distortion of the optical path, and flux calibration relative to standard stars \citep{tubiana-guettler2015}. 

A list of OSIRIS images obtained before, during, and after the outburst is given in Table~\ref{tab:image_parameters}, and the properties of the employed filter bands are listed in Table~\ref{tab:filters}.
We here analyse images obtained during the last months of the Rosetta mission when the spacecraft was close to the comet, providing high spatial resolution.
\begin{table}
\centering
\caption{Observational circumstances characterising the OSIRIS images used for this work. ``C'' defines the camera (NAC/WAC). The local time (LT) is calculated as 12+($\lambda_{site} - \lambda_{sun}$)/15, where $\lambda_{site}$ and $\lambda_{sun}$ are the longitudes of the outburst site and of the subsolar point in degrees. The azimuth (Az) and zenith distance (ZD) of the spacecraft are given in degrees and were calculated with respect to the vector, $r_0$, from the origin of the comet reference frame to the outburst site. Az is the angle between the components perpendicular to $r_0$ of the north direction and the vector, $r_{s/c}$, from the outburst site to the spacecraft. ZD is the angle between $r_0$ and $r_{s/c}$. The last column gives the distance, $D$, between the outburst site and the spacecraft in km. Double horizontal lines separate groups of images obtained under similar circumstances. The pixel scale is given by $a D$, where $a_{NAC}=18.6\,\mu$rad and $a_{WAC}=101\,\mu$rad.}
\label{tab:image_parameters}
\begin{tabular}{lllrrrr}
\hline
Obs. Date [UT] & C & Filter & LT[h] & ZD & Az & $D$\\
\hline
2016-05-03 00:42 & N & or. & 10.9073 &  53.8 &   51.1 & 17.906\\  
2016-07-03 07:50 & W &    red & 11.0239 &  48.1 &   58.9 &  8.534\\  
\hline\hline
2016-07-03 07:04 & W &    red & 9.4979 &  38.3 &   78.0 &  8.557\\  
2016-07-03 08:47 & W &    red & 12.9304 &  68.9 &   51.9 &  8.768\\  
2016-07-03 08:50 & W &    red & 13.0144 &  69.9 &   51.9 &  8.783\\  
\hline\hline
2016-07-02 21:26 & N &   blue & 14.3332 &  34.2 & 116.6 & 11.859\\  
2016-07-02 21:26 & N & or. & 14.3395 &  34.2 & 116.5 & 11.857\\  
2016-07-02 21:26 & N &    NIR & 14.3459 &  34.1 & 116.4 & 11.856\\  
2016-07-02 21:36 & N &   blue & 14.6649 &  32.9 & 111.4 & 11.776\\
2016-07-02 21:36 & N & or. & 14.6712 &  32.9 & 111.3 & 11.775\\
2016-07-02 21:36 & N &    NIR & 14.6776 &  32.9 & 111.2 & 11.773\\
\hline\hline
2016-03-19 21:26 & N & or. & 14.8761 & 43.4 & 145.0 & 10.847\\
2016-03-19 21:46 & N & or. & 15.5388 & 36.8 & 140.3 & 10.746\\
\hline\hline
2016-05-02 12:59 & N & or. & 11.5852 &  54.4 &   81.2 & 17.917\\  
2016-05-02 12:59 & N &   blue & 11.5981 &  54.4 &   81.2 & 17.918\\  
2016-05-02 13:00 & N &    NIR & 11.6276 &  54.5 &   81.1 & 17.920\\  
\hline
2016-07-09 20:47 & W &    red & 11.4596 &  51.7 &  106.7 & 10.317\\  
2016-07-09 21:46 & W &    red & 13.3904 &  50.2 &   90.8 &  9.938\\  
\hline
2016-07-24 10:15 & N &   blue & 11.0542 &  51.3 &   98.0 &  8.505\\  
2016-07-24 10:15 & N & or. & 11.0606 &  51.2 &   97.9 &  8.505\\  
2016-07-24 10:15 & N &    NIR & 11.0670 &  51.2 &   97.9 &  8.504\\  
2016-07-24 10:30 & N &   blue & 11.5519 &  50.8 &   93.5 &  8.483\\  
2016-07-24 10:30 & N & or. & 11.5583 &  50.8 &   93.5 &  8.483\\  
\hline
2016-08-21 13:19 & W &    red & 10.8456 &  43.9 &   77.9 &  7.033\\ 
\hline\hline
2016-01-06 08:51 & N & or. &  8.8523 &  46.2 &   34.9 & 84.652\\
\hline
2016-05-03 12:05 & N & or. &  9.5692 &  46.1 & 17.3 & 17.731\\
2016-05-03 12:25 & N & or. & 10.2322 &  54.9 & 20.0 & 17.899\\
\hline
\end{tabular}
\end{table}
\begin{table}
\centering
\caption{List of employed OSIRIS filters and their properties. $\lambda_c$: Central wavelength; $\Delta \lambda$: bandwidth; $I_{sun}$: solar flux at central wavelength and 1\,AU.}
\label{tab:filters}
\begin{tabular}{lllrrrr}
\hline
Camera & Filter & $\lambda_c$ [nm] & $\Delta \lambda$ [nm] & $I_{sun}$ [W\,m$^{-2}$\,nm$^{-1}$]\\
\hline
NAC & NIR & 882.1 & 65.9 & 0.9230\\
NAC & orange & 649.2 & 84.5 & 1.5650\\
NAC & blue & 480.7 & 74.9 & 2.0300\\
WAC & red & 629.8 & 156.8 & 1.7000\\
\hline
\end{tabular}
\end{table}
The dust plume observed by WAC at 07:50 (Fig.~\ref{fig:outburst_site}) was optically thick and cast a measurable shadow on the surface. It originated between two boulders (B1 and B2). The site emerged from the shadow of the northeastern wall of Basin F at 07:30 (local time 10:17). A detailed analysis of this and additional images is found in Section~\ref{sec:results}.

\subsection{Alice}
\noindent Alice is a far-ultraviolet (70--205~nm) imaging spectrograph onboard  Rosetta that observed emissions from various atomic and molecular species in the coma of comet 67P/Churyumov-Gerasimenko \citep{feldman-ahearn2015} as well as reflected solar radiation from both the nucleus and the dust coma \citep{feaga-protopapa2015}.  Alice employed a two-dimensional photon counting detector that accumulated counts over an interval, usually 5 or 10 min, into a histogram array of wavelength vs. spatial position along the 5.5$^\circ$ slit.

\begin{figure}
\includegraphics[width=\columnwidth]{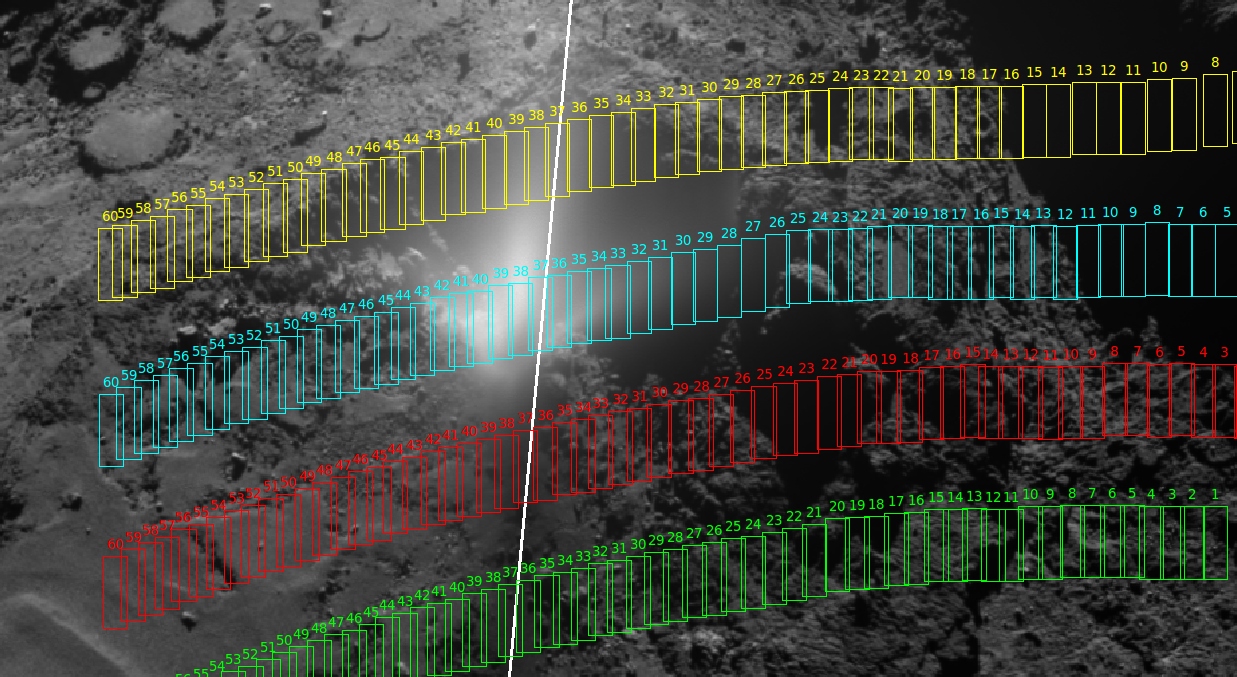}
\caption{Position of the Alice slit overlayed on the WAC image of the outburst (Fig.~\ref{fig:outburst_site}) as a function of time. The rectangles correspond to individual pixels of the Alice detector, each 0.3$^\circ$ $\times$ 0.1$^\circ$, of which rows 4 (green), 6 (red), 8 (blue), and 10 (yellow) are shown.  The numbers indicate minutes after UT 07:00, and the white line marks the position of the slit at UT 07:37. The pixel positions were calculated using the pixel centres relative to the central boresight given in the ROS\_ALICE\_V16.TI instrument kernel in the spice library \citep{acton1996}.}
\label{fig:alice_on_wac}
\end{figure}

\begin{figure}
\includegraphics[width=\columnwidth]{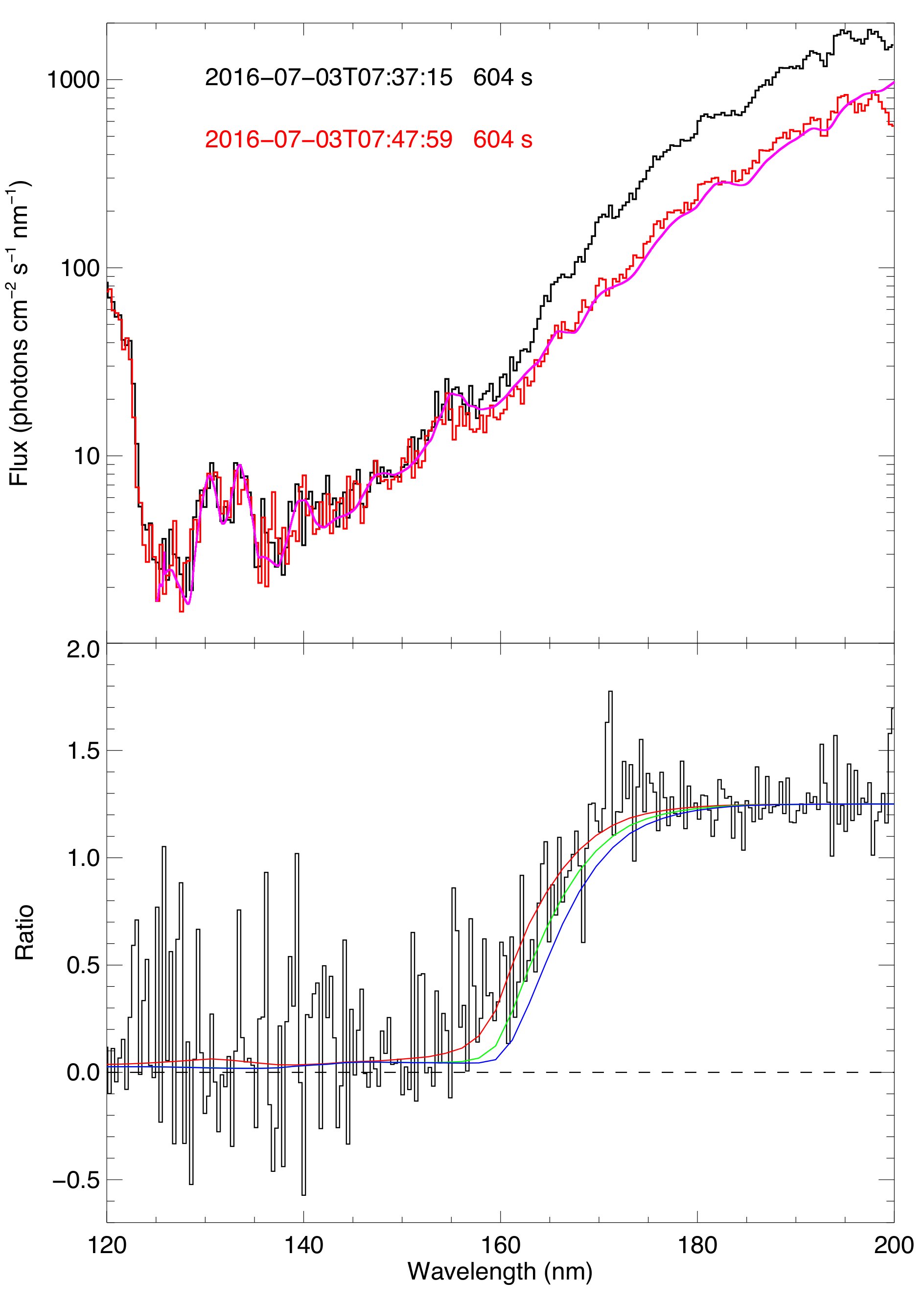}
\caption{The top panel shows two Alice spectra, each a 604 second integration. The black line histogram begins at UTC 07:37:15, and the peak count rate is determined from the OSIRIS image to occur at $\sim$07:39. The red line is the following spectrum beginning at UTC 07:47:59, and does not show the long wavelength enhancement due to the outburst but only solar reflected light from the surface. A scaled solar spectrum \citep{mcclintock-rottman2005}, convolved to the Alice resolution, is shown (in magenta) for comparison. We assume the black spectrum to be that of the outburst grains superimposed on that of the surface, while the red spectrum is surface alone. The difference is then the spectrum of the released grains. The lower panel shows this difference divided by the surface spectrum to give the normalized bidirectional reflectance spectrum. Water ice models with grains of diameter 0.2\,$\mu$m (red); 0.5\,$\mu$m (green); and\,1.0\,$\mu$m (blue), from \citet{hendrix-hansen2008} are shown. These demonstrate that the grains in this particular outburst are composed of sub-micron water ice particles.}
\label{fig:alice_spectrum}
\end{figure}

\noindent At the time of the outburst, Alice was obtaining histograms of 604\,s 
integration time. The histogram beginning at 07:37 shows a very large increase in reflected solar radiation in the wide bottom of the Alice slit. From analysis of the OSIRIS image (Fig.~\ref{fig:alice_on_wac}) we find that the outburst is confined to a region 1.2$^\circ$ $\times$ 0.1$^\circ$, which translates to a projected footprint on the nucleus of 180\,m x 15\,m. Because of the rotation of the comet and the motion of Rosetta, the outburst is seen by the Alice slit for only $\sim$3 minutes of the 10-minute histogram. In comparison to the subsequently obtained spectrum (beginning at 07:47), the spectrum covering the outburst plume shows increased flux at long wavelengths (Fig.~\ref{fig:alice_spectrum}). The sharp absorption edge below 170~nm is characteristic of sub-micron water ice particles \citep{hendrix-hansen2008}. This spectral feature appears only in the histogram beginning at 07:37, and only in the wide bottom of the Alice slit, and thus can be uniquely associated with the plume observed by OSIRIS/WAC 10 minutes later. Grains in the vicinity of the spacecraft are likely too optically thin to be detected against reflected sunlight from the surface, so we cannot determine if they are water ice. Unlike the outbursts of volatile gas observed by Alice on multiple dates around perihelion \citep{feldman-ahearn2016}, no gas emission associated with this event is detected. The Alice housekeeping data, with a time resolution of 30\,s, show a rise in the total count rate beginning from 07:36. While this does not necessarily mark the beginning of the outburst due to the motion of the FOV across the comet surface, it represents the latest possible start time.

\subsection{GIADA}
\label{subsec:giada}
GIADA on board Rosetta was designed to determine the physical properties of cometary dust: momentum, speed, mass and the geometrical cross section of individual particles \citep{dellacorte-rotundi2014,dellacorte-rotundi2016}. The information on single particles was derived by two subsystems mounted in cascade: The Grain Detection System (GDS) and the Impact Sensor (IS). The GDS detected particles crossed a laser curtain providing their cross sections and triggering a time of flight counter from the GDS to the IS. From this time measurement, the particle speed was retrieved. The IS consisted of a sensing plate equipped with five piezoelectric sensors (PZTs). A particle impacting the sensing plate generated bending waves detected by the PZTs. The PZTs signal was monotonically related to the particle momentum. GIADA detections, depending on which sub-system detected the individual dust particle, are divided in: 1) GDS-only detections providing the particle cross-section and speed; 2) IS-only measuring particle momentum; and 3) GDS-IS detection determining individual particle mass, speed and geometrical cross-section. Since the subsystem detections depend on the particle physical characteristics (size, optical properties, density) thanks to different detection types we classified different classes of particles. GDS-only detections, occurring as isolated events or as "dust showers", i.e. up to hundreds of detections in tens of seconds, correspond to ultra-low-density ($<$1\,kg\,m$^{-3}$) aggregates \citep{fulle-dellacorte2015}. IS-only and GDS+IS detections are of compact particles with average density of 800\,kg\,m$^{-3}$ \citep{fulle-dellacorte2016}.

\begin{figure}
\includegraphics[width=\columnwidth]{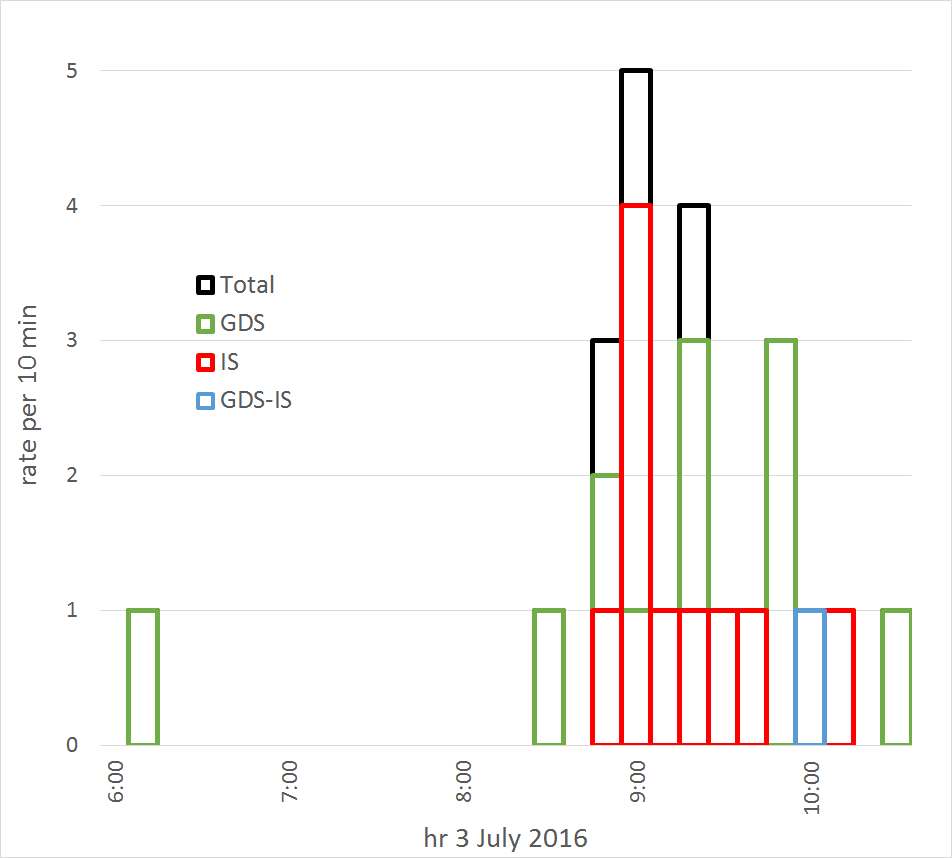}
\caption{Particle detection rate measured by GIADA. The different event types are colour coded. The one GDS+IS detection occured at UT 09:56 and revealed a velocity of (1.37 $\pm$ 0.08)\,m\,s$^{-1}$ and a particle mass of 10$^{-7}$\,kg, corresponding to a 310\,$\mu$m radius sphere of density 800\,kg\,m$^{-3}$.}
\label{fig:giada}
\end{figure}

\begin{figure}
\includegraphics[width=\columnwidth]{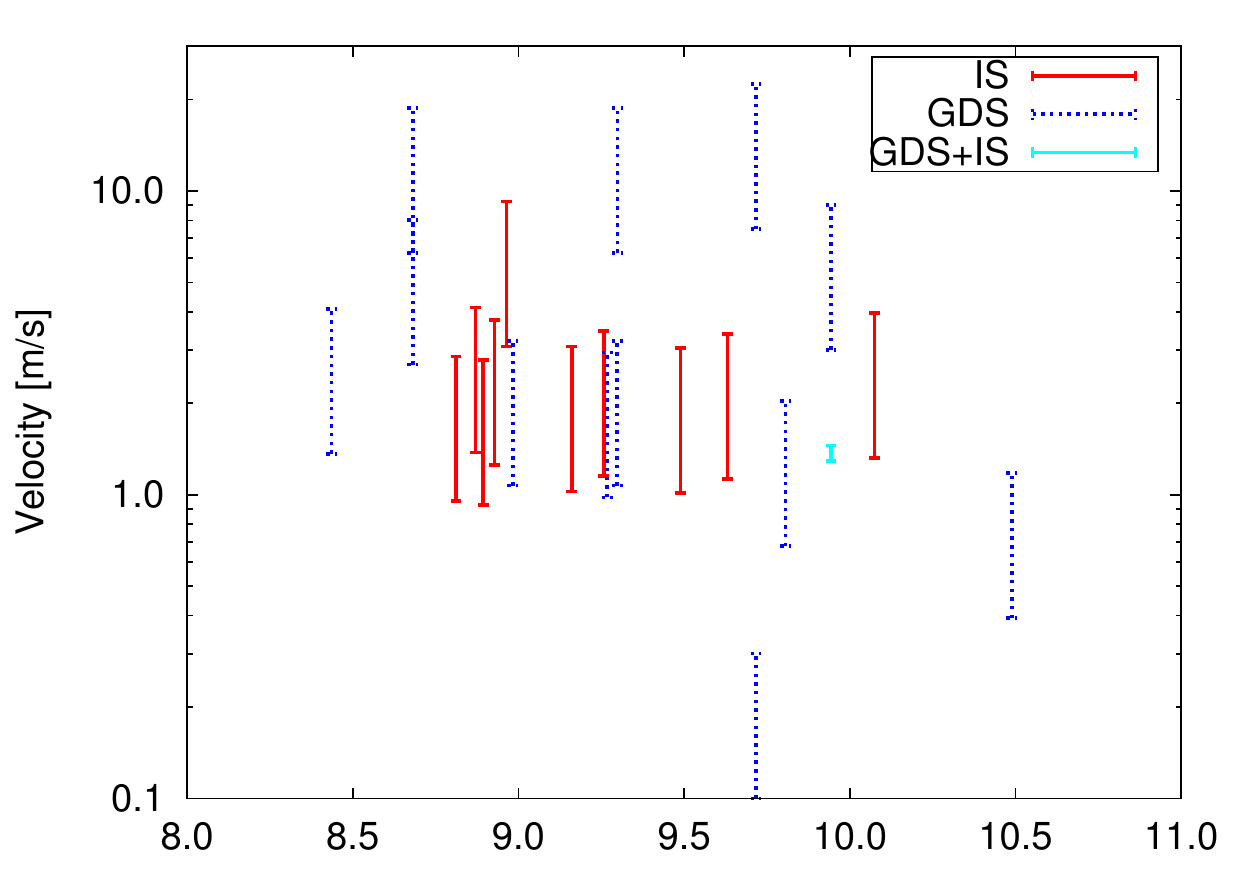}
\includegraphics[width=\columnwidth]{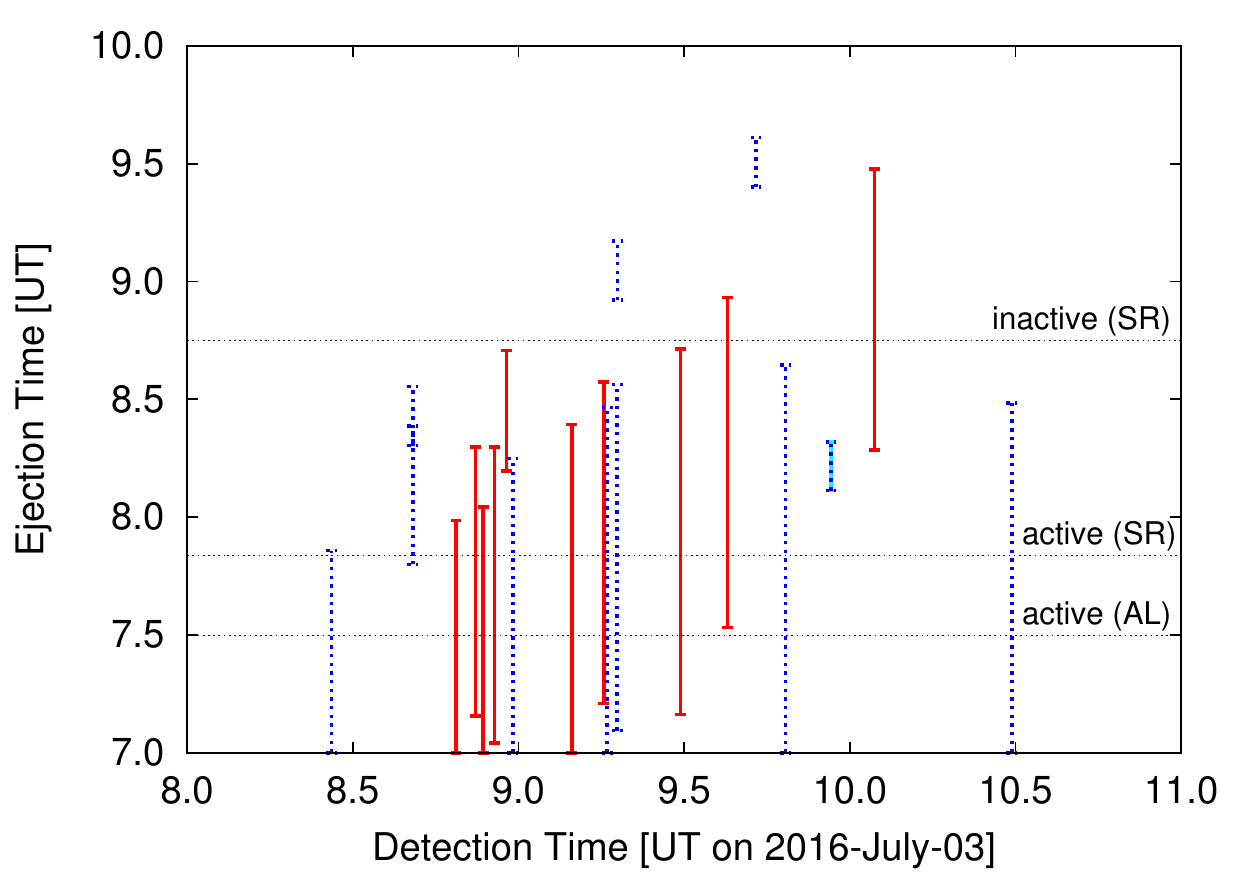}
\caption{Velocities (top panel) and inferred starting times (bottom panel) of particles detected by GIADA. The velocities of the IS-only particles were determined using the method described in \citep{dellacorte-rotundi2016} using a value for the reference speed, $A$, optimized for this particular event such that all IS-only particles are compatible with ejection during the time interval confined by the first detection of activity by Alice and the observation of the inactive surface by OSIRIS. The error bars correspond to a velocity uncertainty of 50\% for the IS-only and GDS-only detections.}
\label{fig:giada_speed}
\end{figure}

GIADA detected 22 particles (1 GDS-IS, 10 IS, and 11 GDS) between UT 8:26:29 and 10:29:21 on July 03 (Fig.~\ref{fig:giada}). This detection peek was embedded in a period with almost no other detections. The total number of events during the week from July 01 to 07 was 25, such that 90\% of the detected material in this time period are associated in time with the outburst. The peak of detections occured around UT 09:00, and the particle speeds were typically $<$3\,m\,s$^{-1}$ (Fig.~\ref{fig:giada_speed}). 50\% of the particles were detected by the IS, suggesting that these were compact with bulk densities of order 800\,kg\,m$^{-3}$ \citep{fulle-dellacorte2016}. It is possible that the detection peak around UT 09:00 and the following decrease in detection rates reflect the temporal evolution of the activity level.

\subsection{COSIMA}
\label{subsec:cosima}
\begin{figure*}
\includegraphics[width=\textwidth]{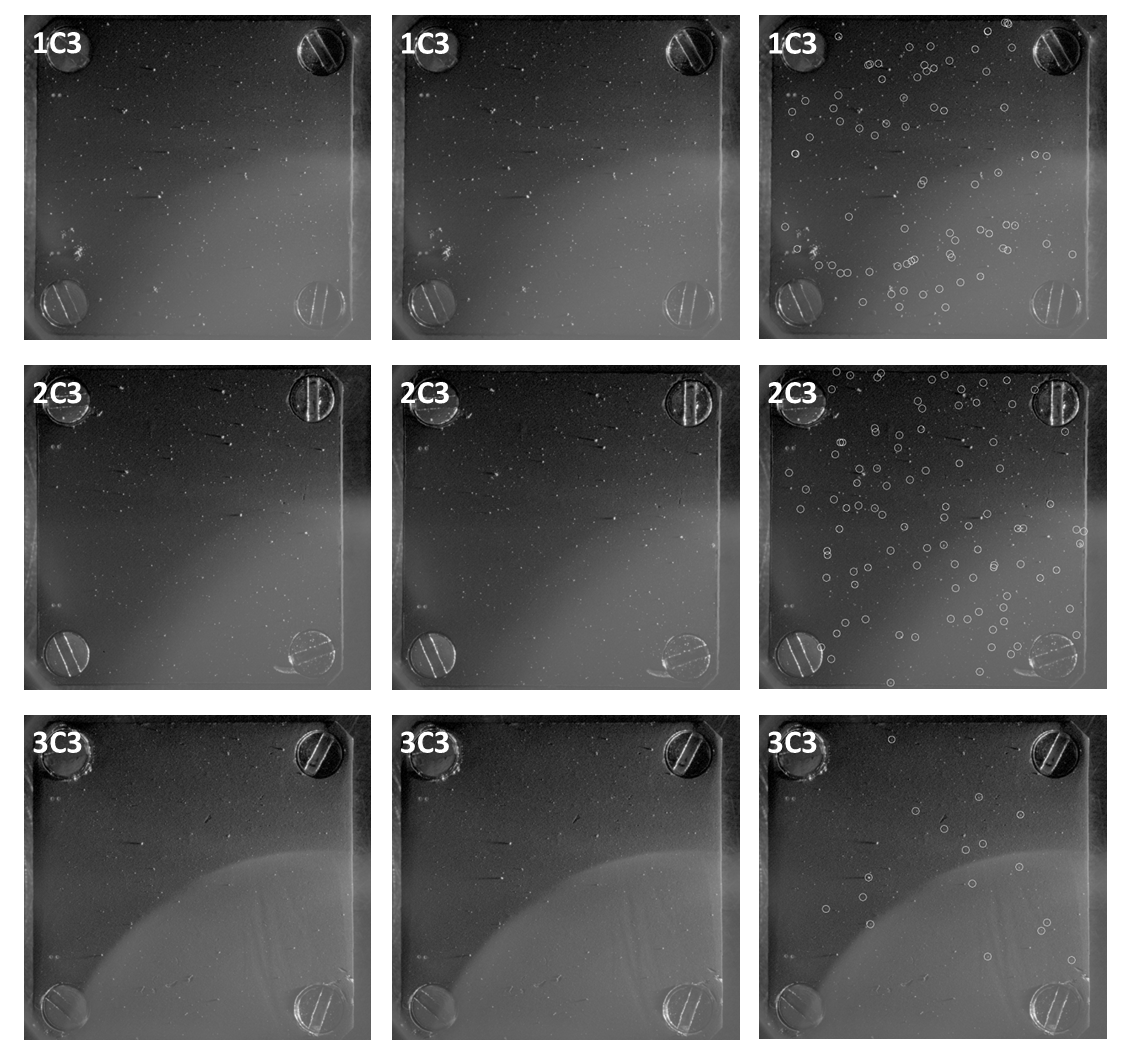}
\caption{COSISCOPE optical microscope images of targets: left column: before the exposure period of 2016 July 01 -- 07, centre column: after the exposure, and right column: with the locations of the particles that were collected during this period marked.}
\label{fig:cosima_targets}
\end{figure*}

COSIMA was a Time-Of-Flight mass spectrometer on the Rosetta orbiter that collected dust particles in the coma of 67P on substrate frames, with 3 mounted metallic targets of 1\,cm$^2$ each. With an optical microscope camera, COSISCOPE, the dust particles were imaged with a resolution of 14\,$\mu$m $\times$ 14\,$\mu$m \citep{kissel-altwegg2007,langevin-hilchenbach2016}.
The three metallic targets were exposed at the same time to the cometary dust flux. The typical exposure periods ranged from a few hours up to 3 weeks. Images of the target holder were acquired prior to and after each exposure period. New particles were identified by comparison of the two image sets. The target holder was located at the rear of a 14.9\,cm-long funnel with a FOV of 15$^\circ \times$ 23$^\circ$. The cometary particles passed through the funnel before impact on the targets (Fig.~\ref{fig:cosima_targets}). The particles collected by COSIMA were able to fragment at very low velocity \citep{hornung-merouane2016}. Some particles hit the funnel walls prior to impact on the target and likely broke into pieces. These pieces, if they did not stick to the funnel, were scattered and eventually stuck on a target. In order to estimate the number of the primary dust particles that entered the instrument's FOV, this scattering and fragmentation has to be taken into account. To distinguish between individual incoming particles and pieces of particles created by the fragmentation of a larger parent particle, the spatial distribution of the particles collected for each exposure period is analyzed. If the particles are not randomly spatially distributed, they are assumed to come from the fragmentation of a unique parent particle inside the funnel. The method to determine which particles result from such events is described in \citet{merouane-zaprudin2016}.

During the exposure periods preceding as well as following the first week of July, very few particles have been collected (see Table~\ref{tab:cosima}). During the week of the outburst, 188 particles were detected on the targets. However, the analysis of their spatial distribution shows that they are likely to be the fragments of a single large parent particle that disintegrated in the funnel. GIADA detected 22 particles connected in time to the outburst event, with only 2 other particles within a week's interval around July 3, and COSIMA was hit by one particle that fragmented on the target. Given that the total collection areas of GIADA and COSIMA differ by a factor of 30, the total numbers of detected particles are consistent, and the COSIMA particle stems with high probability from the outburst.

If we sum the volumes of all the fragments on the COSIMA targets, assuming they have a half-sphere shape, accounting for some flattening upon impact, we obtain a total volume of 2.3$\times$10$^{-12}$\,m$^3$. This volume corresponds to a spherical particle of about 160\,$\mu$m in diameter. As fragments from the parent impacting particle may have stuck to the dust funnel's walls, the particle size stated is a lower limit estimate.

\begin{figure}
\includegraphics[width=\columnwidth]{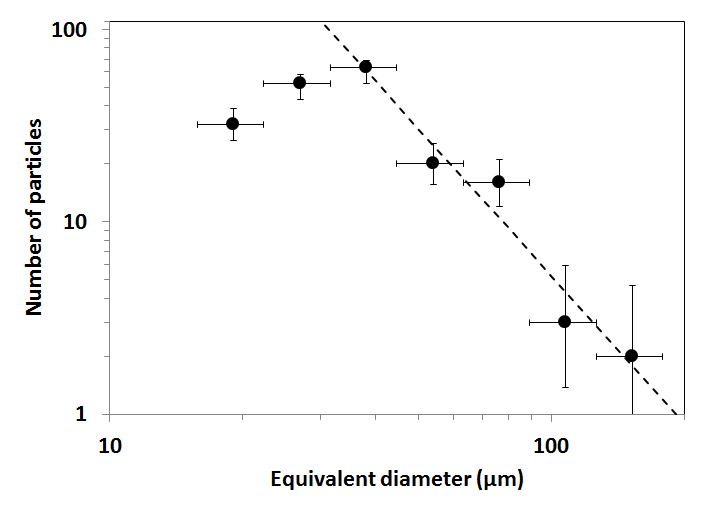}
\caption{Size distribution of particle fragments collected on the COSIMA targets during the exposure period including the outburst. The cumulative power index is -2.54, which is close to the value measured for all the particles collected after perihelion (-2.58). However, since the particles from this outburst result from the fragmentation of a single parent broken in the funnel, the size distribution reflects that of the fragments. The measured power index should rather be compared to the value of -2.3 $\pm$ 0.2 measured for the size distribution of fragments of particles \citep{hornung-merouane2016}. The method used to determine the number of particles and the error bars is described in \citet{merouane-zaprudin2016}.}
\label{fig:sd_cosima}
\end{figure}

The size distribution of the fragments of the collected particle gives a hint on the tensile strength and/or the velocity of the incoming particle. A very fragile or very fast particle would tend to break into a lot of small fragments whereas a low velocity and/or strongly bounded aggregate would tend to break into few big pieces. The cumulative power index of the size distribution shown in Fig.~\ref{fig:sd_cosima} is -2.54. This power index is very close to the average power index of -2.3 $\pm$ 0.2 measured for the size distribution of fragments of particles collected before perihelion at heliocentric distances ranging from 3.57 to 2.36 AU \citep{hornung-merouane2016}. This implies that the particle tensile strength is of the same order of magnitude than previously reported values of several hundreds of Pa. Then the velocity would range from 2\,m\,s$^{-1}$ to 5\,m\,s$^{-1}$, and the particle density would be 200-300 kg\,m$^{-3}$ \citep{hornung-merouane2016}. This density range implies that the estimated mass of the particle collected during this week is (0.5 - 0.8) $\times$ 10$^{-9}$ kg. With an assumed density of 1000\,kg\,m$^{-3}$, the total mass would be higher, 2.3$\times$10$^{-9}$ kg, and would imply a higher tensile strength and higher velocity of the impacting particle \citep{hornung-merouane2016,merouane-zaprudin2016}.

\begin{table}
\centering
\caption{Numbers of particles collected by COSIMA during the period between 2016 June 06 and July 19. Between June 08 and July 01, no target was exposed. The first two columns list the start and end date (in 2016) of the exposure period. $N_{f}$ is the number of individual particle fragments identified on the target, and $N_{p}$ is the number of parent particles from which these fragments have been inferred to derive from. The quantity $N_{p}$ is assumed to be 1 if the spatial distribution of the particles collected during the exposure period of interest is not random. The method to determine $N_{p}$ is described in \citet{merouane-zaprudin2016}. The volume, $V$, is calculated by deriving the radius of the particles on the target assuming a circular shape and calculating the volume of a half-sphere (to take into account a flattening of the particles after impact on the target).}
\label{tab:cosima}
\begin{tabular}{llrrr}
\hline
$T_{start}$ [UT] & $T_{end}$ [UT] & $N_{f}$ & $N_{p}$ & $V$ [m$^3$]\\
\hline
Jun-06 03:04:19 & Jun-06 06:40:13 & 0   & 0 & 0 \\
Jun-06 08:24:45 & Jun-07 13:15:14 & 0   & 0 & 0 \\
Jun-07 14:55:26 & Jun-08 00:00:11 & 1   & 1 & 8.3$\times$10$^{-15}$\\
Jul-01 12:52:09 & Jul-07 07:13:45 & 188 & 1 & 2.3$\times$10$^{-12}$\\
Jul-07 10:40:55 & Jul-13 05:55:10 & 19  & 1 & 1.5$\times$10$^{-12}$\\
Jul-13 09:22:12 & Jul-16 20:41:30 & 11  & 1 & 1.4$\times$10$^{-13}$\\
Jul-17 00:08:54 & Jul-19 19:03:37 & 5   & 5 & 3.8$\times$10$^{-13}$\\
\hline
\end{tabular}
\end{table}

\subsection{STR-B}
\label{subsec:str}
The Rosetta spacecraft carried two identical Star Trackers (STR-A and STR-B) as part of its attitude control system \citep{buemi-landi2000}. The STR cameras had apertures of 29\,mm, an effective focal length of 46\,mm, and a FOV of 16.4$^{\circ}\times$16.4$^{\circ}$. They were equipped with CCD detectors comprising 1024 $\times$ 1024 pixels, and their sensitivity extended over a broad range in the visible spectrum.

In the nominal tracking mode, the instrument continuously measured the position and magnitude of up to 9 stars in the FOV in order to derive the spacecraft attitude. In this process the background signal was determined in 20 $\times$ 20 pixel windows containing the tracked stars. The average of this quantity over all tracking windows is available as a housekeeping parameter downlinked with a typical sampling interval of 32\,s. The parameter value was then bias-corrected and converted into spectral radiance units based on information extracted from the magnitude calibration relations applied by the instrument for stellar targets.

\begin{figure}
\centering
\includegraphics[width=\columnwidth]{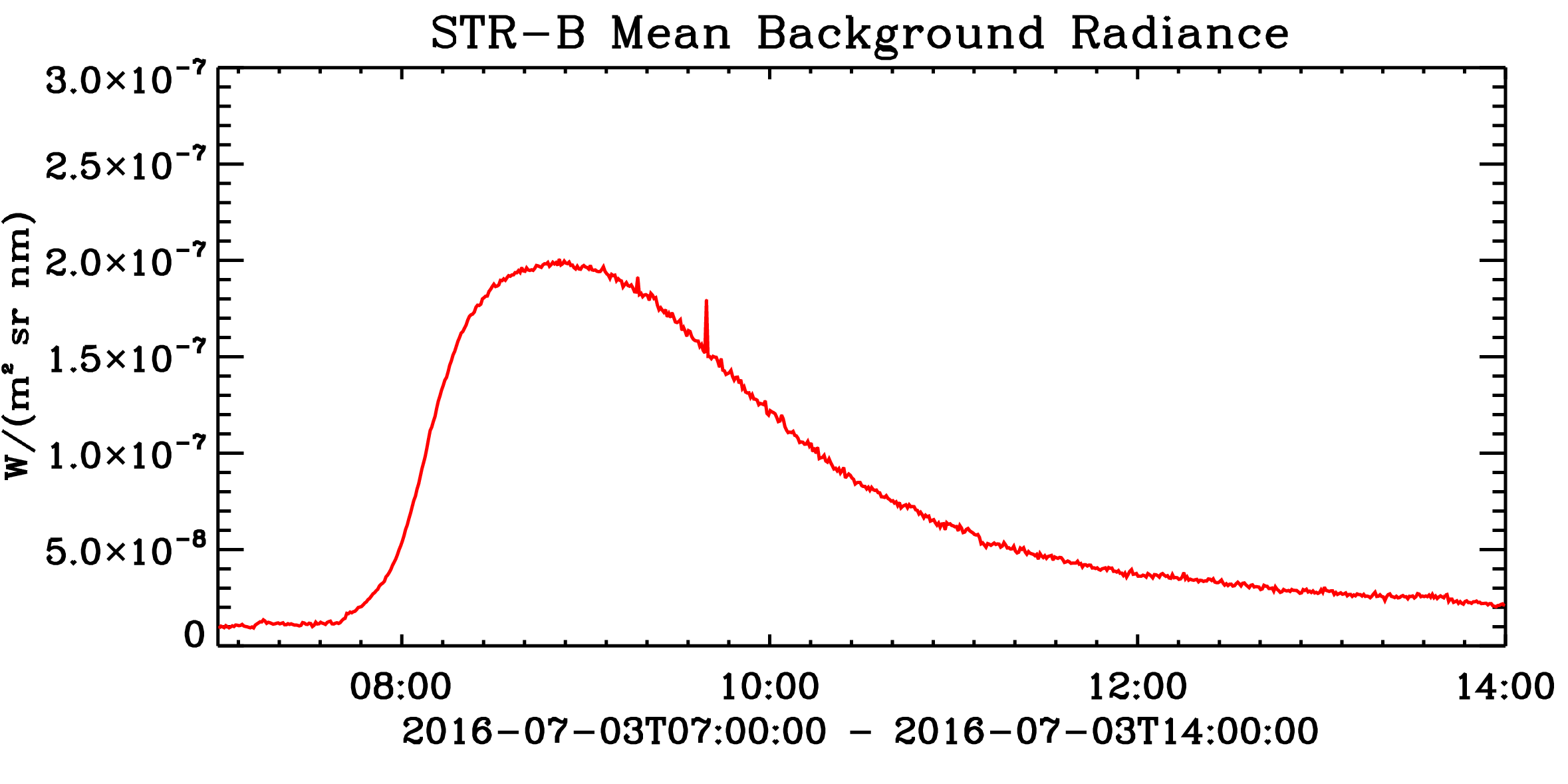}
\caption{Surface brightness in the FOV of the STR-B as a function of time.}
\label{fig:flux_STR}
\end{figure}

During the time of the outburst, STR-B was continuously obtaining data, measuring the average surface brightness, $I_{STR}$ in its FOV. Mounted on the spacecraft with a boresight offset by 100$^\circ$ from that of the science instruments and pointing away from the main outburst site, the STR-B would have detected dust from the outburst only when it arrived close to the spacecraft, such that it can be directly set in relation to the measurements by the in situ instruments.
Fig.~\ref{fig:flux_STR} shows the temporal profile of the surface brightness measured by the STR-B. The peak flux was observed at UT 08:50, about simultaneously with GIADA. This suggests that the dominant optical cross-section of dust released during the outburst was in particles in the sensitivity range of GIADA, a few 100\,$\mu$m in radius. 

\section{Results} 
\label{sec:results}

\subsection{Dust velocity}
\label{subsec:velocity}
Assuming that the outburst started between 7:30 (first illumination) and 7:36 (first Alice detection), the fastest particles, detected by the STR-B at 07:40, travelled at a speed of (25 $\pm$ 10) m s$^{-1}$. Particles seen by STR-B at 14:00 and having started from the comet after 7:30 and having left the OSIRIS FOV by 8:48 had a velocity of (0.41 $\pm$ 0.05)\,m\,s$^{-1}$. 

\subsection{Outburst duration and timing}
\label{subsec:duration}
Assuming that the outburst did not start before the site was exposed to sunlight at 7:30 and that the slowest particles (0.41\,m\,s$^{-1}$) would have needed 10 minutes to leave the OSIRIS FOV (250 pixel distance) by 8:48, we limit the duration of the outburst to a maximum of 68 minutes. 

If the outburst had been an instantaneous event, it would need to have occured not later than at 7:36. In that case, even the slowest known particles travelling at 0.41\,m\,s$^{-1}$ would have been at 350\,m distance from the outburst site at 7:50, and all particles visible in the OSIRIS image would have been slower than those detected by STR-B at 14:00. In that case, Alice would have observed much faster particles than OSIRIS, likely those creating the in situ detection peak between 8:30 and 9:00, and -- based on the STR-B measurement -- the surface brightness measured by Alice should have been a factor 10 higher than that measured by OSIRIS, which is in contrast to the observed value of 0.25 (Section~\ref{subsec:surf_brightness}). We therefore exclude an instantaneous event and assume that the activity was still on-going at 7:50, giving a minimum duration of 14 minutes.

\subsection{Plume surface brightness and shadow}
\label{subsec:surf_brightness}

\begin{figure}
\centering
\includegraphics[width=\columnwidth]{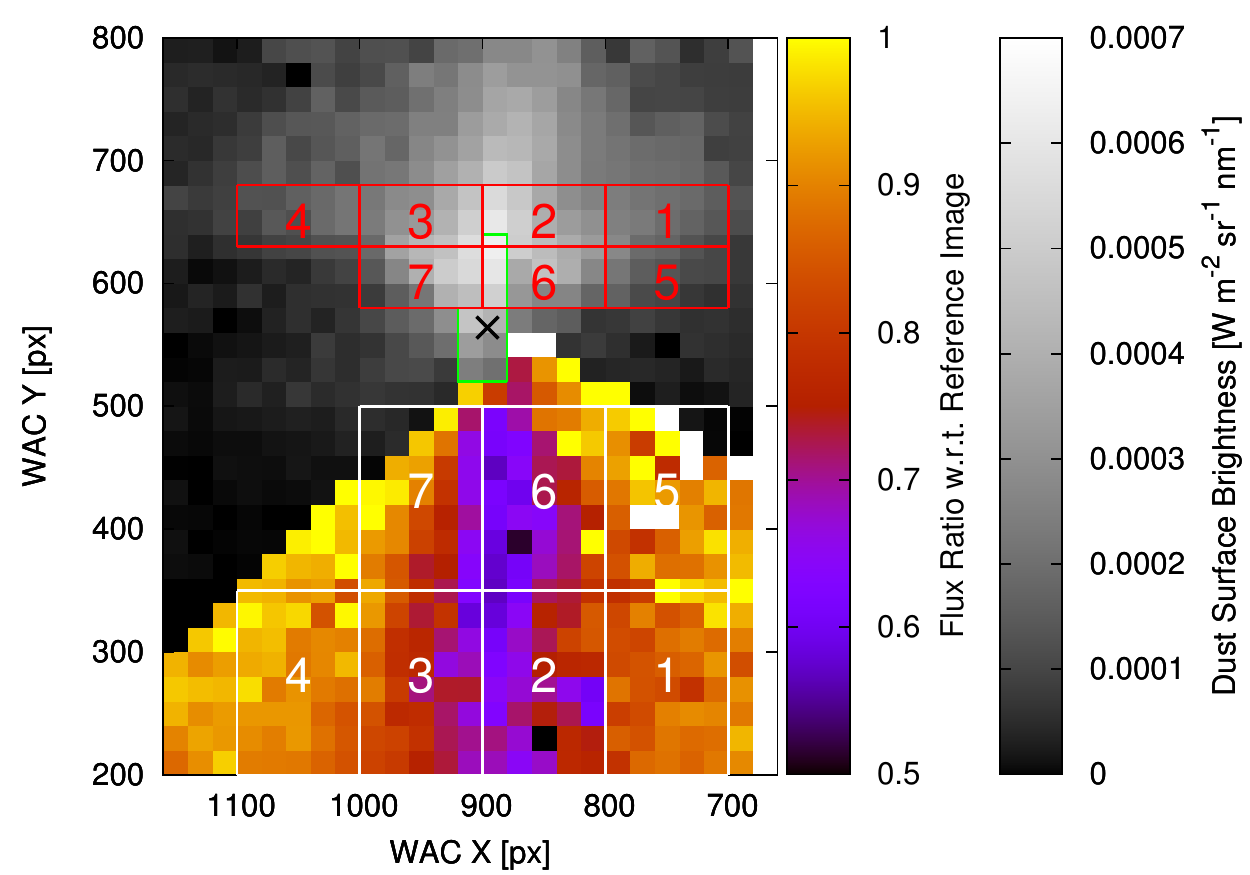}
\caption{The outburst image subtracted (greyscale) or divided (colour) by the May 03 image scaled to account for the different heliocentric distances and filter bands. The greyscale represents the surface brightness of light scattered by dust in the plume, while the colour code indicates the fraction of light prevented from reaching the surface by the dust. In the central region (framed in green), the surface brightness is given as on July 03, assuming that the dust column was optically thick. Each square has a projected linear size of 17.2\,m at the distance of the comet. The numbered fields were used to derive the dust albedo in Sec.~\ref{subsec:albedo} from the extinction of light reaching the surface (white fields) and the surface brightness of light scattered by dust in the plume (red fields). The geometric association of the fields is based on the assumption that the darkest region of the shadow was cast by the brightest section of the plume.}
\label{fig:albedo_fields}
\end{figure}

To evaluate the amount of light (1) scattered by dust in the plume towards the camera and (2) prevented from reaching the surface, we use the NAC image of 2016 May 03 00:42 (Fig.~\ref{fig:outburst_site}, right) as a reference for the brightness of the surface in the absence of a plume. In both images, we measured the surface brightness in $\sim$3600 circular apertures having projected radii of 1.72\,m. The positions of these apertures were manually selected to match the same landmarks such as boulders and their shadows in both images, resulting in $\sim$3600 pairwise measurements of the surface brightness, each pair consisting of one measurement on the outburst image, $I^j_{outburst}$, and one on the reference image, $I^j_{ref}$, were $j$ identifies the pair. The surface brightness measured on the reference image was scaled by a factor $f$ = 0.842 to account for the different heliocentric distance and central wavelength. For each pair, we calculated the ratio $R = I_{outburst}/(f I_{ref})$ (colour-coded in Fig.~\ref{fig:albedo_fields}) and the difference $I_{plume} = I_{outburst} - f I_{ref}$ (greyscale in Fig.~\ref{fig:albedo_fields}), and averaged over all pairs within 20x20 pixel squares of the outburst image. To calculate the average value of $R$ in a square, we considered only illuminated spots on the surface. The ratio $R$ characterizes the depth of the shadow and is meaningful in the predominantly shadowed region of the surface. The difference, $I_{plume}$, corresponds to the amount of light scattered by dust in the plume. In the central region of the plume, no contours of the underlying cometary surface could be identified even at highest possible stretching of the brightness scale. We interpret this as an optically thick region of the plume and assume that all light received was scattered or absorbed by dust. The shadowed region closest to the plume origin (around coordinates (900,500) in Fig.~\ref{fig:albedo_fields}) is seen through a considerable amount of foreground plume dust that increases the value of $R$. In general, the innermost region of the plume (860$<x<$920, 500$<y<$600) seems to be characterized by a complex interplay of light scattering by the dust and by the surface and shadowing by the dust. An interpretation of this region would require a detailed model of the dust distribution and light scattering and is beyond the scope of this paper. It is possible that the lowest part of the plume appears relatively dark because it is shadowed by the dust above.

The typical plume surface brightness in the WAC red filter in the background-subtracted OSIRIS image is of order 4$\times$10$^{-4}$ W m$^{-2}$ sr$^{-1}$ nm$^{-1}$, corresponding to a radiance factor $I/F$ of 8$\times$10$^{-3}$. Alice measured a surface brightness of about 4000 rayleighs in the spectral range of 175 -- 195 nm, corresponding to $I/F$ = 2$\times$10$^{-3}$. Simultaneous measurements of OSIRIS and Alice on 2016 February 19 found a ratio of 2 between the OSIRIS- and Alice-measured radiance factors \citep{gruen-agarwal2016}, as opposed to a factor of 4 in the present data. This slight difference is explained by the motion of the Alice slit during the 10 minute integration, such that the central plume was covered by the slit only for about 30\% of the total integration time.

\subsection{Plume orientation and shape}
\label{subsec:plume_geometry}
The plume in Fig.~\ref{fig:outburst_site} has an opening angle of (100 $\pm$ 5)$^\circ$, possibly with a denser central region of (25 $\pm$ 5)$^\circ$ opening angle. Associating the brightest region in the plume $P0$=(890,630) in Fig.~\ref{fig:albedo_fields} with the darkest part of the shadow $S0$=(890,370), we calculate the 3-dimensional point of intersection, $PI$, of the lines connecting the point of intersection of $S0$ with the surface to the Sun, and $P0$ to the spacecraft using the shape model SHAP5 v1.5 \citep{jorda-gaskell2016} and the SPICE toolkit \citep{acton1996}. We define the footpoint of the plume, $PF$, from the intersection of the edges of the inner and outer cone, and calculate its 3-dimentional position from the intersection of the line of sight crossing $PF$ with the shape model. We interpret the line connecting $PF$ and $PI$ as the central axis of the cone. Had the plume originated from the centre of the comet reference frame, its direction would have corresponded to latitude -41$^\circ$ and longitude 158$^\circ$ with an uncertainty of 10$^\circ$ resulting from the size of the 20$\times$20 pixel blocks used for the triangulation, and an additional, difficult to quantify, uncertainty arising from the uncertainty of the pixel association. At 7:50, the line connecting $PF$ to Rosetta was at an angle of (35 $\pm$ 10)$^\circ$ from the central plume axis. This angle increased over the following hours.

\subsection{Dust albedo}
\label{subsec:albedo}
We constrain the product of albedo and phase function at a phase angle of 95$^\circ$, $P$, of the plume material from the OSIRIS image in two ways. (1) Assuming that the innermost region of the plume was optically thick, we derive a lower limit of $P_{centre}$=0.012. This is one order of magnitude brighter than the nucleus with $P_{nucleus}$ = 1.1$\times10^{-3}$ \citep{fornasier-hasselmann2015}. 

In the second approach, we derive $P$ from the cross section of dust casting the shadow on the surface and the brightness of light scattered by dust. For each of the 20$\times$20 squares shown at colour-scale in Fig.~\ref{fig:albedo_fields}, we calculated the projected area perpendicular to the solar direction $A_{\perp}$, and assumed that the total geometric cross section of dust between the square, $i$, and the Sun was given by $C_i$=(1-$R_i$) $A^i_\perp$, where $R_i$ is the fraction of light removed from the incident flux in square $i$ (Sec.~\ref{subsec:surf_brightness}). We then grouped the squares to seven larger fields of 100$\times$150 pixels (white boxes in Fig~\ref{fig:albedo_fields}), and associated each of these fields with a complementary field in the coma (red boxes), again assuming that the darkest part of the shadow was caused by the brightest region of the plume (Section~\ref{subsec:plume_geometry}).
We obtain for each pair of fields $j=[1,7]$ the total dust cross section $C_j$ in m$^{2}$ and the intensity of the scattered light $I_j$ in W\,m$^{-2}$\,nm$^{-1}$, and calculate the product of geometric albedo and phase function, $P_{plume}^j$, from

\begin{equation}
P_{plume}^j = \frac {I_j}{C_j} \frac{\pi r_h^2 \Delta^2}{I_{sun}},
\end{equation}
where $r_h$ is the heliocentric distance in AU, $\Delta$ is the distance to the spacecraft in m, and $I_{sun} = 1.7$\,W\,m$^{-2}$\,nm$^{-1}$ is the solar flux at 1\,AU in the WAC/red bandpass.
We find $P_{plume}$ = 0.025 $\pm$ 0.002, where the given uncertainty corresponds to the standard error of the mean from averaging over all seven fields. The similarity of values obtained for the different fields seems to support the chosen association of fields on the ground and on the plume.
The measured value of $P_{plume}$ is a factor of $\sim$20 brighter than that of the nucleus. This factor is comparable to the ratio of a typical cometary dust phase function $\phi_{dust} (\alpha=95^\circ)$ = 0.5 \citep{kolokolova-hanner-cometsII-2004}, and the phase function of the nucleus $\phi_{nucl} (\alpha=95^\circ)$ = 0.02 \citep{fornasier-hasselmann2015}. 

A possible interpretation is that the plume consisted of a mixture of dark refractory grains of several hundred micrometers having a nucleus-like albedo and phase function, and a component of bright ice grains as detected by Alice. The phase function of ice crystals depends strongly on their shape and texture.
For 0.1 $< P_{ice} <$ 0.2 ~\citep{liu-mishchenko2006}, an admixture of (12 - 25)\% in cross-section of such bright grains would be required to raise the average by a factor 20. \citet{liu-mishchenko2006} model the geometric albedo of ice grains of various shapes as a function of the phase angle. At $\alpha=95^\circ$, they find $P_{ice}$ = 0.1 for spheres and $P_{ice}$ = 0.2 for a mix of cylinders, while the value for spheroids is between these extremes.
Their computations were done for a size distribution of particles with the efficient particle radius of 5\,$\mu$m and at a wavelength of 1.88\,$\mu$m. Since in light scattering the
defining characteristic is the ratio of radius to wavelength, then 5/1.88
should give the same results as particles of radius 1.5\,$\mu$m at the
wavelength 0.6\,$\mu$m used in our observation, which is slightly larger than the particle size derived by Alice. The imaginary part of the refractive index of ice is slightly larger at 1.88\,$\mu$m (1.e-4) than at visible wavelengths (1.e-6). But this difference in absorption is not significant enough to strongly affect the results.

\subsection{Dust mass and production rate}
\label{subsec:mass}

The total dust cross section ($C=\sum_i C_i$) causing the observable shadow is 6200\,m$^{2}$. If (75 - 88)\% of this cross-section were contributed by particles of a representative 250\,$\mu$m radius, the corresponding volume would be (1.6 - 1.8)\,m$^3$, or a disk of 10\,m radius and (5 - 6)\,mm thickness. For a density in the range $\rho$=(250 .. 795)\,kg\,m$^{-3}$, the equivalent mass would be $M_{image}$ = (920 $\pm$ 530)\,kg. The main uncertainty of the mass estimate results from the uncertainty of the average density, the range of which we chose to include both the COSIMA value of 250\,kg\,m$^{-3}$ (Sec.~\ref{subsec:cosima}) and the GIADA average density of 795\,kg\,m$^{-3}$ \citep{fulle-dellacorte2016}.
The mass $M_{image}$ is a lower limit to the total ejected dust mass, because an unknown fraction of the dust was outside the FOV, and because the activity continued over a time interval between 14 and 68 minutes. 

The extinction cross section was measured on dust at a distance of 20 to 100\,m from the source (Fig.~\ref{fig:albedo_fields}). At these distances, the dust travels at (90 -- 105)\% of the terminal velocity \citep{jewitt_133P}. 
Assuming a representative dust velocity of 2\,m\,s$^{-1}$ (Sec.~\ref{subsec:giada}), dust leaves the shadow-casting region of the plume in $\sim$50\,s. This gives a dust production rate of $Q_M$ = (18.4 $\pm$ 10.6)\,kg\,s$^{-1}$ or a number flux of $Q_N$ = (5.1 $\pm$ 0.4)$\times$10$^{8}$ s$^{-1}$. For a duration of 14 -- 68 minutes and assuming a constant production rate, the estimated total mass of ejected dust is between 6\,500 and 118\,000\,kg. The excavated volume lies in the range (26 -- 150)\,m$^3$, corresponding to a layer of (8 -- 47)\,cm for a 10\,m radius patch.

If the dust was distributed uniformly across a cone with an opening angle of 100$^\circ$ (solid angle of 2.24\,sr), we expect a number flux of $F_{GD}$=(0.0025 $\pm$ 0.0002)\,s$^{-1}$ on the 0.01\,m$^2$ sensitive area of GIADA (subtending a solid angle of 1.1$\times$10$^{-11}$\,sr as seen from the comet). This leads us to expect a detection of (23 $\pm$ 2) particles over a 2.5\,h time period, in good agreement with the 22 actually detected particles.\\

\subsection{Dust size distribution}
\begin{figure}
\centering
\includegraphics[width=\columnwidth]{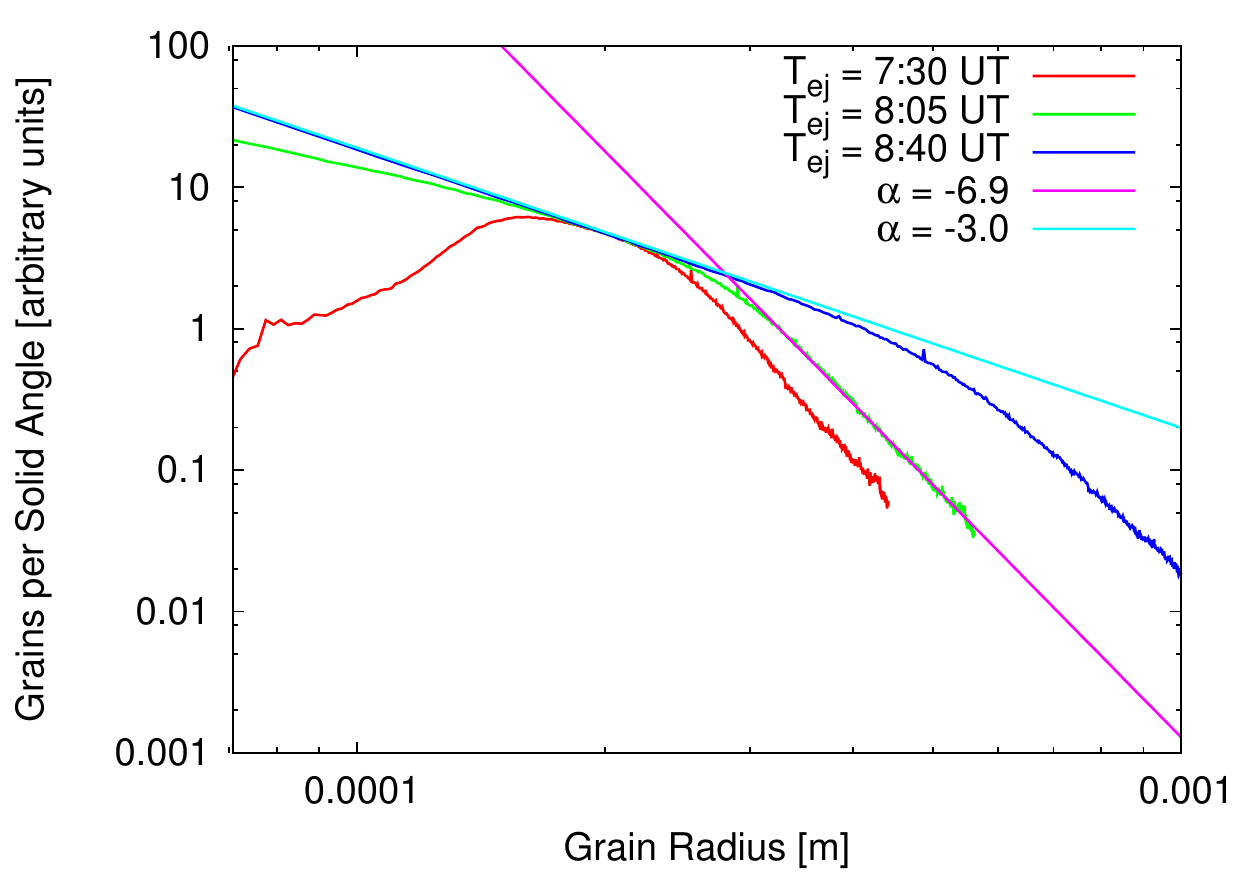}
\caption{Relative number of particles in the FOV $N = I_{STR} / s^2$ plotted vs. $s = s_0 \sqrt{t/t_0}$ for three different values of the starting time $t_{start}$. The data follow power laws of different exponents for  150$\mu$m $< s < s_t$ and for $s > s_t$ . For $s <$ 150\,$\mu$m (fast particles), the error introduced by the unknown starting time is too large to infer the dust size distribution, while this uncertainty only moderately affects the transition size (230$\mu$m $< s_t <$ 500\,$\mu$m) and not the exponents.}
\label{fig:sd_STR}
\end{figure}

\noindent In the following, we derive the dust size distribution from the combined measurements of STR-B and GIADA. If the outburst had been of negligible duration, we could associate a single particle size passing by the spacecraft to any given time. Assuming that the relative distribution of the spatial dust density along the STR-B line of sight and the dust optical properties did not change with time, then $(I_{STR} - I_{bg}) / s^2$ reflects the relative number of particles of size $s(t)$ in the FOV, where $I_{bg}$ = 10$^{-8}$\,W\,m$^{-2}$\,nm$^{-1}$\,sr$^{-1}$ is the pre-outburst background level, and $t = t_{obs} - t_{start}$ is the time required by the particle leaving the comet at $t_{start}$ to cover the distance $D$ = 8500\,m from the outburst site to the spacecraft, and $v(t) = D/t$. Following \citet{wallis1982}, we assume a unique relation $s \propto v^{-2}$, between particle size, $s$, and velocity, $v$, and hence $s(t) = s_0 \sqrt{t/t_0}$, with $s_0$ = 200\,$\mu$m, and $t_0$ the difference between UT 08:50 and $t_{start}$, for which we use a range of fixed values covering the possible duration of the outburst. The surface brightness $I_{STR}$ is proportional to the number of particles in the FOV, $N$, and their squared radius $s^2$. For a power-law differential size distribution 
\begin{equation}
\frac{dn}{ds} = s^\alpha,
\end{equation}
follows $N(s) \propto s^{\alpha+1}$.
Fitting a broken power-law to the surface brightness as a function of time, we derive the exponents of the differential dust size distribution in Fig.~\ref{fig:sd_STR}. We find a significant change in the exponent, $\alpha$, of the size distribution at a transition size, $s_{t}$, between 230 and 500\,$\mu$m, depending on the assumed starting time. The exponent changes from $\alpha$ = -3.0 for 150$\mu$m $< s <$ $s_t$ to $\alpha$ = -6.9 for $s > s_t$ . The transition size range dominates the optical cross section of the dust. 

\subsection{Outburst site and surface change}
\label{subsec:site}
OSIRIS observed the outburst site in the Imhotep Basin F multiple times during the mission. We here analyse a set of images obtained between January and August of 2016 to characterise the terrain at the outburst site. Selection criteria to assemble the data set included resolution and viewing geometry (Table~\ref{tab:image_parameters}). Some observations consist of multiple exposures in different bandpasses obtained within a short time interval, characterising the spectral properties of the light scattered by the surface.  

\begin{figure*}
\includegraphics[width=\textwidth]{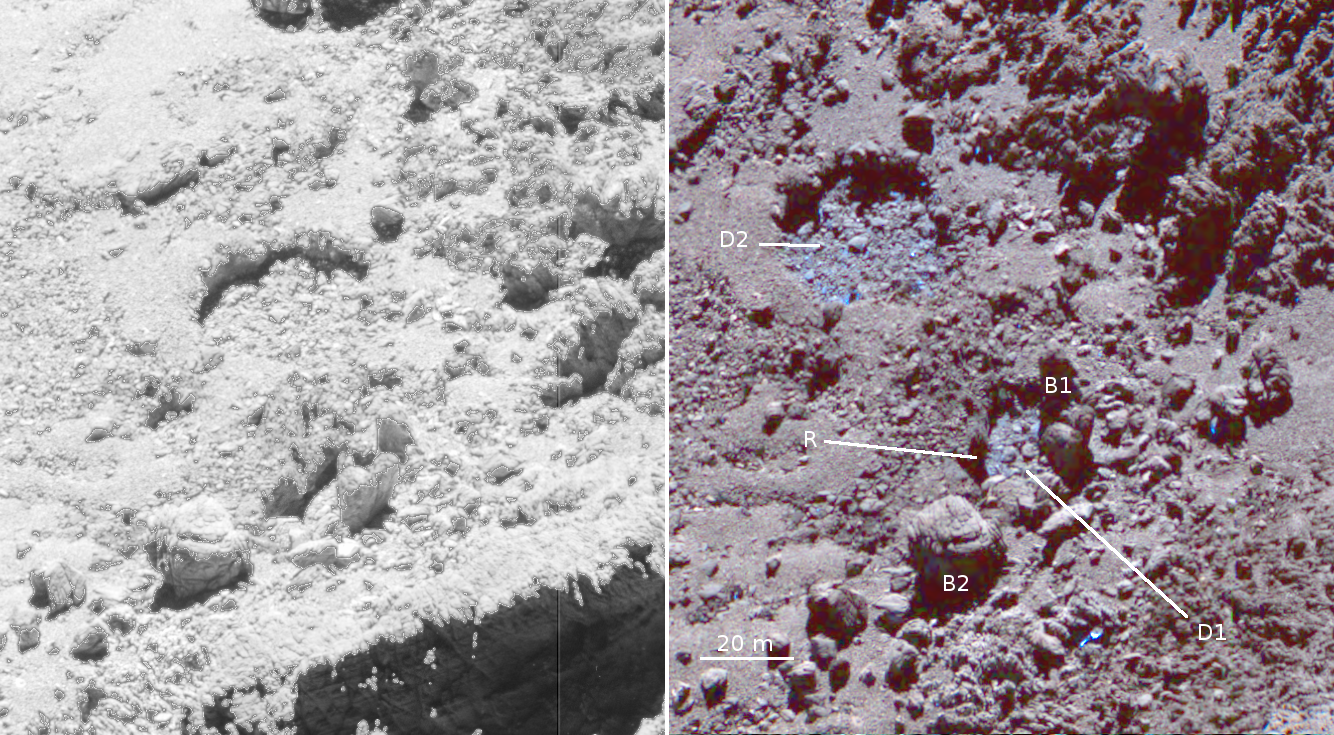}
\caption{The outburst site on 2016 March 19 (left) and July 02 (right). The left panel has two brightness scales: illuminated areas are shown at a logarithmic scale between 0 and 0.001\,W\,m$^{-2}$\,nm$^{-1}$\,sr$^{-1}$, while a linear scale between 0 and 4$\times$10$^{-5}$\,W\,m$^{-2}$\,nm$^{-1}$\,sr$^{-1}$ is employed for areas hidden from direct sunlight but indirectly illuminated by the surrounding bright surface. The right panel is a false-colour rgb composite with the red, green, and blue channels corresponding to the NIR, orange, and blue filters.
The labels in the right panel indicate the roundish depressions D1 at the foot of the plume of 2016 July 03, and D2, close to which a plume was observed on 2016 January 06. The wall R is located where after the outburst a bright blue patch was observed (Fig~\ref{fig:july24_details}), and the boulders B1 and B2 are indicated for orientation. }
\label{fig:mar19-jul02}
\end{figure*}
The plume originated from a 20\,m-diameter, roundish depression (D1) bounded to the southwest by a steep wall, R, of a few meters height and $\sim$20\,m length, to the northeast by a row of larger boulders (B1) and to the southeast by a single large and roundish boulder, B2 (Fig.~\ref{fig:outburst_site},\ref{fig:mar19-jul02}). 

The best images we found of the face of the wall R were obtained on 2016 March 19 with NAC at a resolution of 20\,cm per pixel and lossless compression (Fig.~\ref{fig:mar19-jul02}, left panel). The wall is in shadow, but indirectly illuminated by the surrounding sunlit surface. Its face shows some indications of thermal fracturing similar to that seen on the neighbouring boulders B1 and B2. It is possible that R is overhanging. In that case, the floor below would receive even less sunlight than the wall itself. The dark shadow at the foot of R could hide a crack.

The images constituting the right panel of Fig.~\ref{fig:mar19-jul02} were obtained 10\,h before the outburst. The floor of D1 and of the neighbouring depression D2 shows patches of a bluish colour, indicating the presence of water ice \citep{pommerol-thomas2015,oklay-sunshine2016,barucci-filacchione2016,fornasier-mottola2016}. Since the July 02 observation was obtained at 14:20 local time, and the region emerged from shadow at 10:17 local time, the face of the wall R was illuminated for not more than 4 local hours per rotation in July 2016. 

\begin{figure*}
\includegraphics[width=\textwidth]{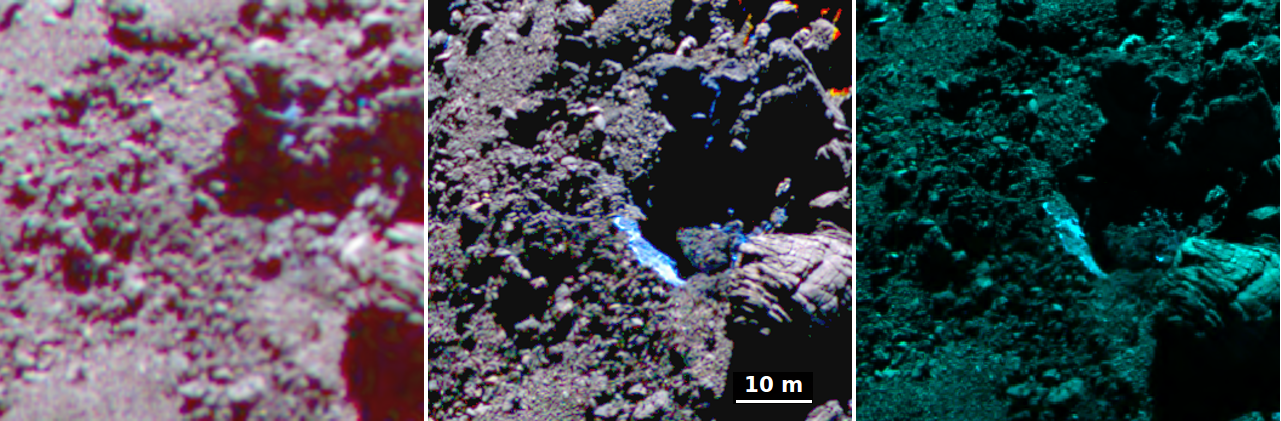}\\[0.05\baselineskip]
\includegraphics[width=\textwidth]{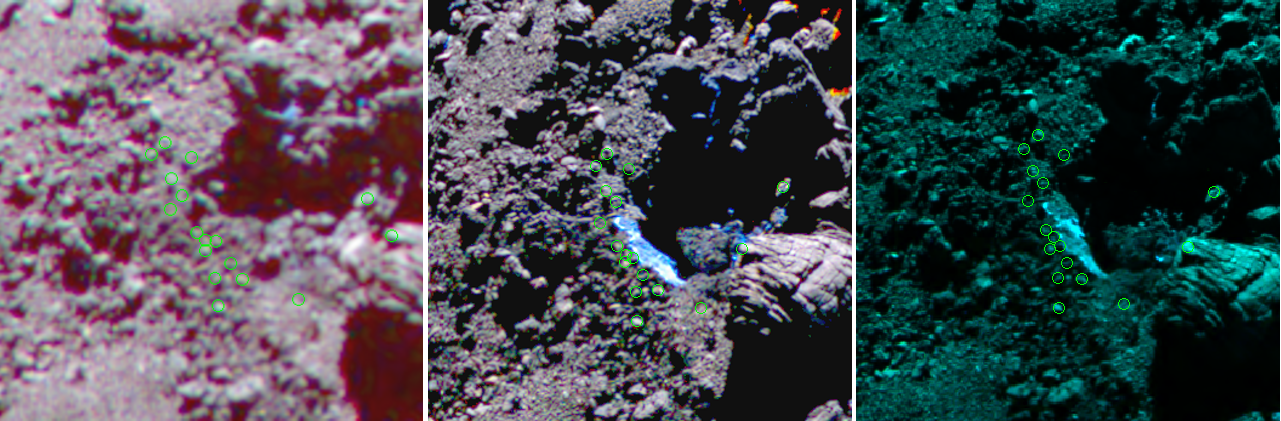}\\[0.1\baselineskip]
\includegraphics[width=\textwidth]{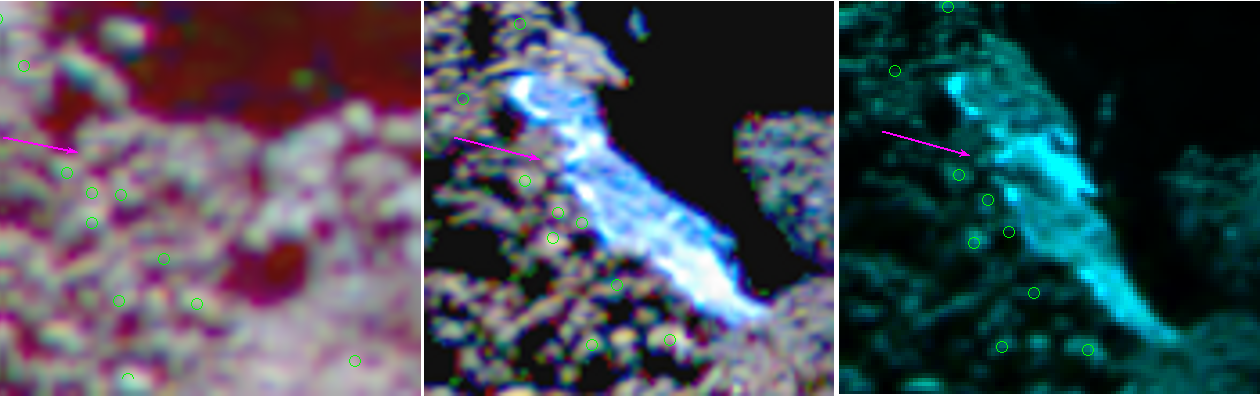}\caption{Left and centre columns: False colour rgb images of the outburst site, with different stretching levels, composed from OSIRIS/NAC images in NIR (red channel), orange (green channel) and blue (blue channel) filters. Right: two-colour composite from orange (green channel) and blue (blue channel) filters, as no NIR data are available. The left column shows the image of May 02, the central and right columns show the images of July 24, 10:15 and 10:30, all at a similar spatial scale (cf. Fig.~\ref{fig:ice_patch} in SM). In the central row, the positions of characteristic surface features are marked by green circles indicating the maximum extent of the region affected by the surface change. 
The bottom row shows a close-up of the bright patch, with the same green circles as above. The arrows point to a boulder row that seems to smoothly continue between the bright area and the surrounding darker area, and to have been in a similar position before and after the outburst. This may indicate that an ice layer froze out on top of the pre-existing surface.}
\label{fig:july24_details}
\end{figure*}

On July 24, NAC images show a bright, bluish patch at the outburst site of a projected size of 15\,$\times$\,5\,m$^{2}$ (Fig.~\ref{fig:july24_details}). The bright patch is also visible in images obtained under similar circumstances on July 09 and August 21 (Fig~\ref{fig:ice_patch}). An image of May 02 (12:59) that was also obtained under similar viewing conditions does not show the bright patch. Comparison of surface features in the May 02 and July 24 images shows that the terrain has not been altered outside a region between the bright patch and the boulder row B1, limiting the size of the affected area to a radius of $\sim$10\,m. 

The visible bright patch is bounded towards the side of B1 by a shadow that gives the impression of a deep crack. To the opposite side, the bright surface makes an abrupt transition to terrain of more typical colour and brightness. The bright patch shows intrinsic brightness variations similar to those of the adjacent surface, which may indicate a similar, boulder dominated terrain. An apparent continuity of the light-shadow pattern between the bright and the typical terrain (marked by the arrow in the bottom panels of Fig.~\ref{fig:july24_details}) may suggest that the bright material is a sharply bounded ice coating on the bouldered surface. Alternatively, the bright material could be a freshly excavated stretch of sub-surface material, and the sharp transition to the typical terrain could be a topographic feature such as the upper edge of a cliff. 

\begin{figure}
\centering
\includegraphics[width=\columnwidth]{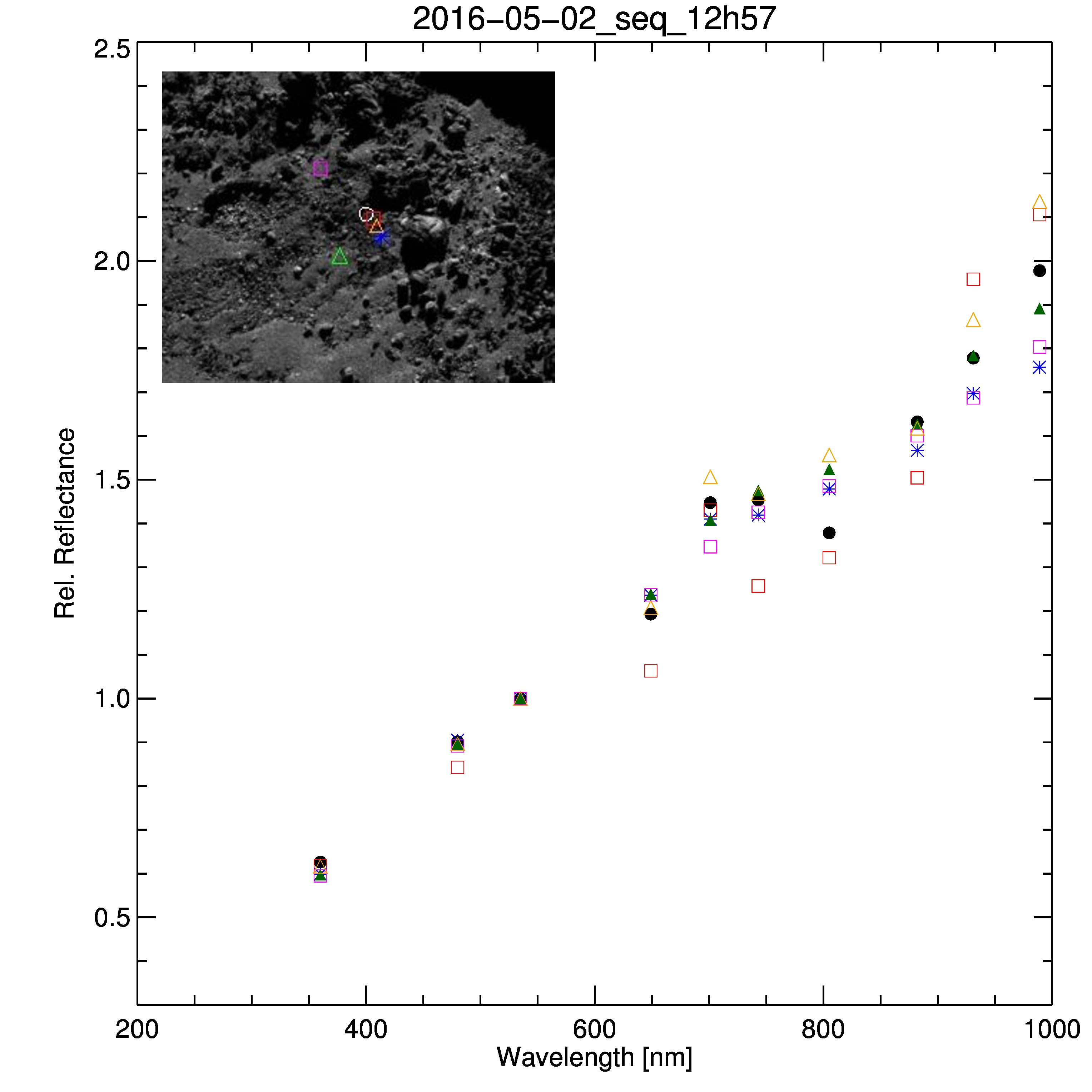}
\includegraphics[width=\columnwidth]{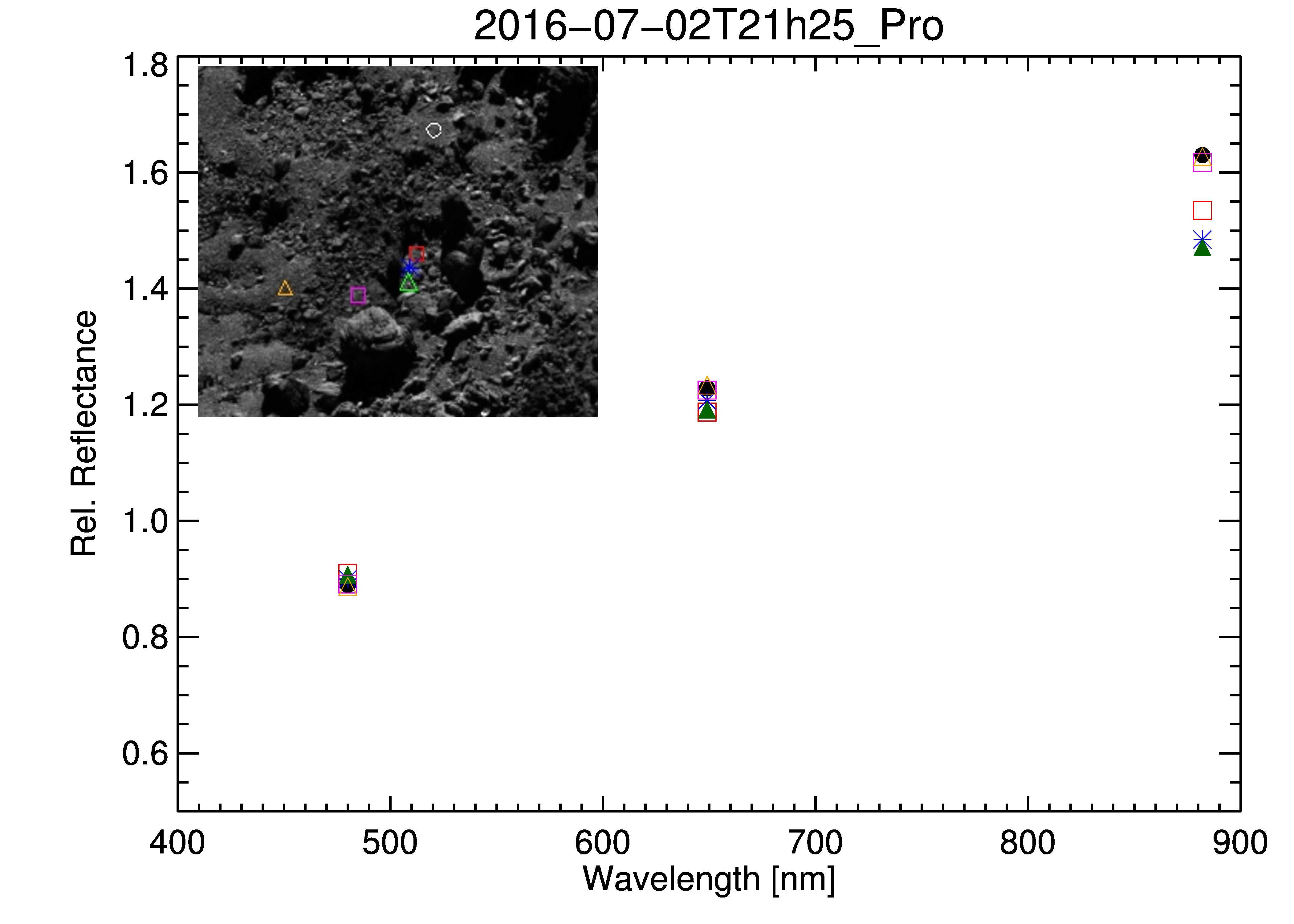}
\includegraphics[width=\columnwidth]{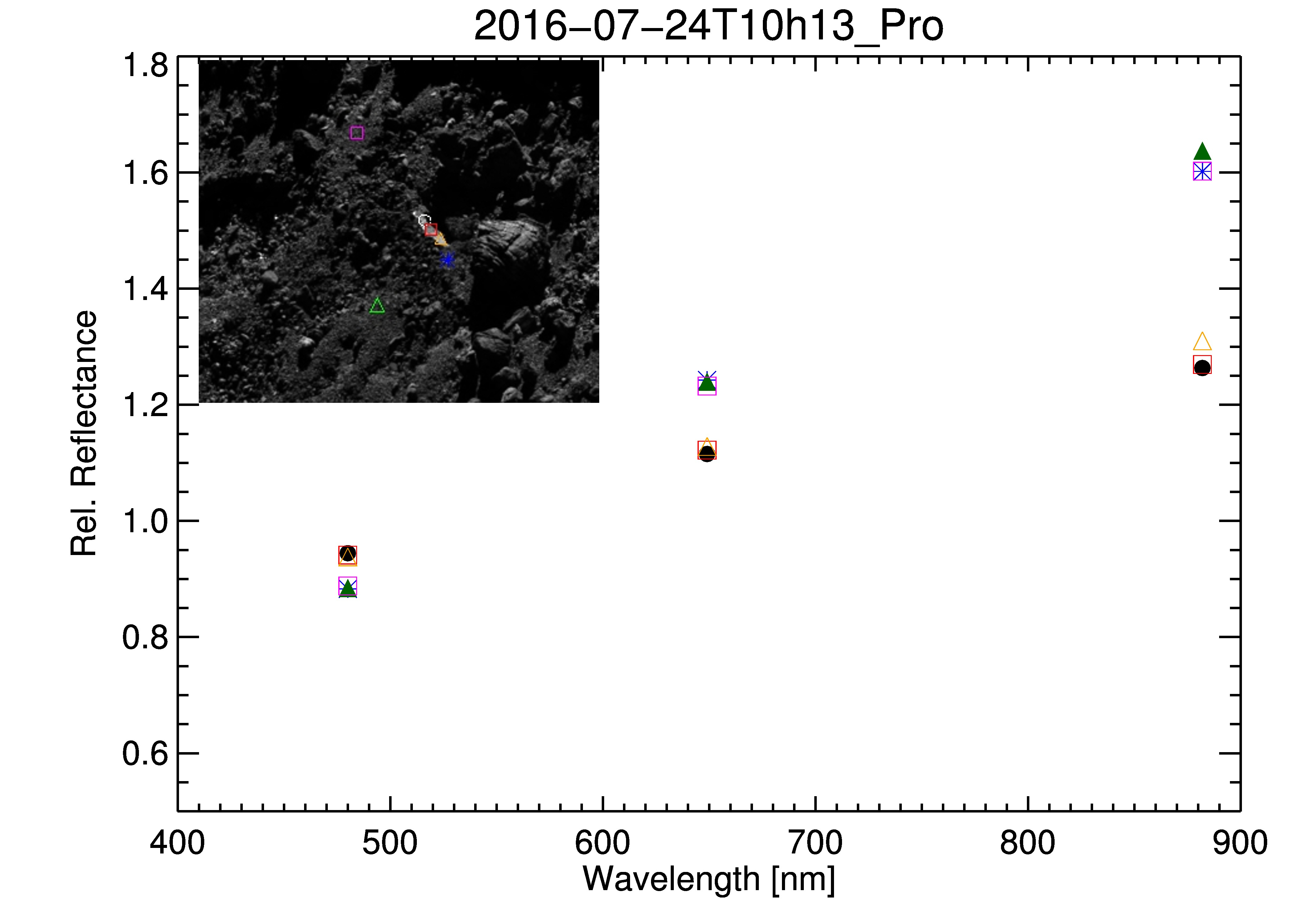}
\caption{Spectral reflectance as a function of wavelength for the six regions of interest following the method described in \citep{fornasier-hasselmann2015,fornasier-mottola2016}. The bright patch on July 24 shows a strong blue colour (red square, black circle, and orange triangle symbols).}
\label{fig:colour_ratios}
\end{figure}

Fig.~\ref{fig:colour_ratios} shows the relative reflectance of selected regions of interests (ROIs), normalised at a wavelength of 535 nm using a linear interpolation between 480 and 649 nm when the observations were not acquired in the green filter centred at 535 nm. Three of the selected ROIs are at the location of the bright patch, and three are on the adjacent terrain. The ROIs in the 24 July bright patch clearly show a much lower spectral slope than the surroundings indicating exposure of some water ice \citep{fornasier-mottola2016, barucci-filacchione2016}.

The bright patch at the outburst site was observable for at least 7 weeks. There are no suitable observations after August 21 to judge on its presence. From the observations of July 02 and 09, we constrain the diurnal duration of solar illumination of the icy spot to 1.18 -- 3.11 local hours, corresponding to 0.59 -- 1.56 hours on Earth. An ice surface in vacuum at the temperature $T$ sublimates at the rate of 
\begin{equation}
Q_{H_2O} = p_{subl}(T) \sqrt{\frac{m_{H_2O}}{2 \pi k_B T}},
\label{eq:QM}
\end{equation}
where $m_{H_2O}$ is the molecular mass of water, $Q_{H_2O}$ is in kg\,s$^{-1}$\,m$^{-2}$, and the sublimation pressure is given by $p_{subl}(T) = A \exp({-B/T})$ with $A$ = 3.56$\times$10$^{12}$\,Pa and $B$ = 6141\,K \citep{fanale-salvail1984}.
We calculate the temperature from the balance of radiative heating and cooling and sublimation cooling:
\begin{equation}
\frac{L}{N_A m_{H_2O}} Q_{H_2O} + \epsilon \sigma T^4 = (1 - A_B) \frac{I_\odot}{r_h^2} \cos \theta + I_{indirect},
\end{equation}
where $\epsilon$ and $A_B$ are the emissivity and Bond albedo of the surface, $\sigma$ and $N_A$ are the Stefan-Boltzmann and Avogadro constants, $L$ = 51000 J/mol is the latent heat of water ice, $\theta$ is the angle between the surface normal and the solar direction, and $I_{indirect}$ represents illumination by scattered light and thermal radiation from other parts of the surface. We approximate this indirect illumination by the expression
\begin{equation}
I_{indirect} = \left[ \frac{I_\odot}{r_h^2} A_B + \epsilon \sigma T_{extern}^4 \right] f_{sky},
\end{equation}
where $f_{sky}$ is the fraction of sky of the primary surface occupied by other parts of the surface having the temperature $T_{ext}$. 

For the emissivity we consider a range between 1 (corresponding to a blackbody) and an extreme of 0.6, consistent with excess temperatures of material in the debris trail \citep{sykes-walker1992a}. 
The average Bond albedo of the 67P surface in the green filter is $A_B$ = 0.012 \citep{fornasier-hasselmann2015}, and we use $A_B$ = 0.24 measured on the Occator bright spots on Ceres \citep{li-reddy2016} as an upper limit for the bright icy patch.

Without indirect illumination, an ice surface having $A_B$=0.012 and $\epsilon=0.6$ illuminated at normal incidence at 3.32\,AU would have a temperature of $T_{max}$=187\,K. For higher albedo and emissivity, and more shallow incidence, the temperature drops to $T_{low}$=177\,K (cf.~Table~\ref{tab:temperatures}). Indirect illumination from a sky fraction of $f_{sky}$ = 0.5 and a dry and therefore hot (215\,K) surface increases the temperatures by a few Kelvin.
Sub-mm and mm-measurements by the MIRO instrument between 3.45 and 3.27\,AU in-bound do not show evidence for near-surface temperatures above 180\,K \citep{schloerb-keihm2015}. 
Sublimation rates corresponding to the above temperature range lie between 4$\times$10$^{-6}$ and 4$\times$10$^{-5}$\,kg\,s$^{-1}$\,m$^{-2}$. 
Assuming a bulk density of 500\,kg\,m$^{-3}$, and an illumination duration of 0.59 -- 1.56\,h per 12.055\,h rotation, we expect that the ice layer eroded by 0.8 -- 28\,mm during 7 weeks, such that its initial thickness must have been at least 1\,mm. 

\begin{table}
\centering
\caption{Equilibrium temperatures of a sublimating ice surface at 3.32\,AU as a function of the Bond Albedo, $A_B$, the emissivity, $\epsilon$, and the incidence angle of sunlight, $\theta$, measured from the zenith. In all except the last line, no indirect illumination was considered. In the last line it was assumed that the surface was additionally heated by scattered visible light and thermal radiation from nearby surfaces at 215\,K covering 50\% of its sky. $^{(1)}A_B$=0.012 was found for comet 67P \citep{fornasier-hasselmann2015}. $^{(2)}\epsilon=0.6$ was derived for the debris trail of comet 67P \citep{sykes-walker1992a}. $^{(3)}A_B$=0.24 was found for Occator bright spots on Ceres \citep{li-reddy2016}.}
\label{tab:temperatures}
\begin{tabular}{llr|rr}
\hline
$A_B$ & $\epsilon$ & $\theta$ & $T$ [K] & Q$_{H_2O}$ [kg\,s$^{-1}$\,m$^{-2}$]\\
\hline
0.012$^{(1)}$ & 0.6$^{(2)}$ & 0$^\circ$ & 187 & 2.64$\times$10$^{-5}$\\
0.012 & 0.6 & 45$^\circ$ & 184 & 1.56$\times$10$^{-5}$\\
0.012 & 1.0 & 0$^\circ$ & 185 & 1.86$\times$10$^{-5}$\\
0.012 & 1.0 & 45$^\circ$ & 181 & 9.05$\times$10$^{-6}$\\ 
0.24$^{(3)}$ & 0.6 & 0$^\circ$ & 185  & 1.86$\times$10$^{-5}$\\
0.24 & 0.6 & 45$^\circ$ & 182 & 1.09$\times$10$^{-5}$\\
0.24 & 1.0 & 0$^\circ$ & 182 & 1.09$\times$10$^{-5}$\\
0.24 & 1.0 & 45$^\circ$ & 177 & 4.25$\times$10$^{-6}$ \\
\multicolumn{3}{l|}{Measured (Ref. \citep{schloerb-keihm2015})}&180 & 3.76$\times$10$^{-6}$ \\
0.012 & 1.0 & 0$^\circ$+ & 189 & 3.72$\times$10$^{-5}$\\
\hline
\end{tabular}
\end{table}

\subsection{Frequency of outbursts}
\label{subsec:jan06}
\begin{figure*}
\includegraphics[width=\textwidth]{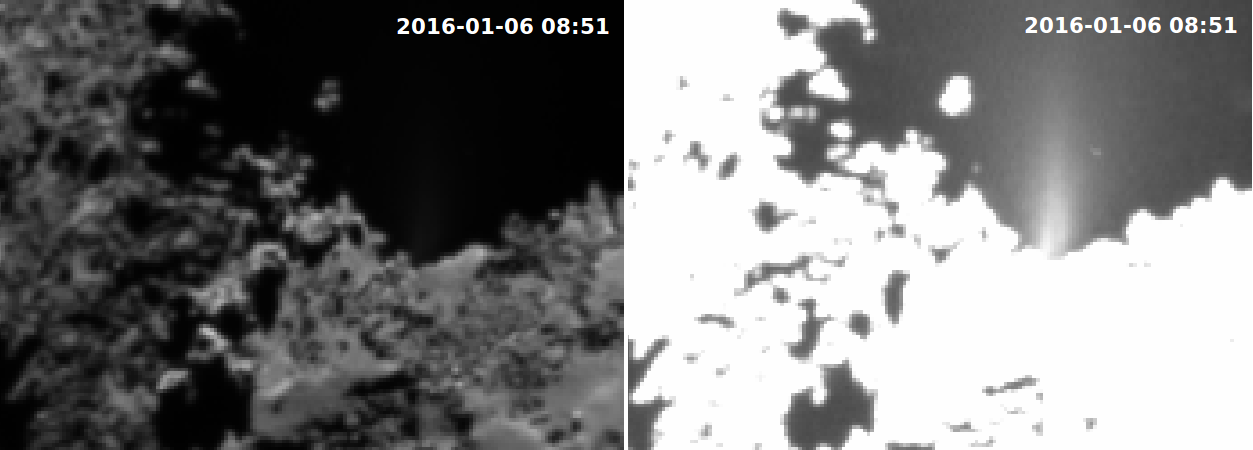}
\includegraphics[width=\textwidth]{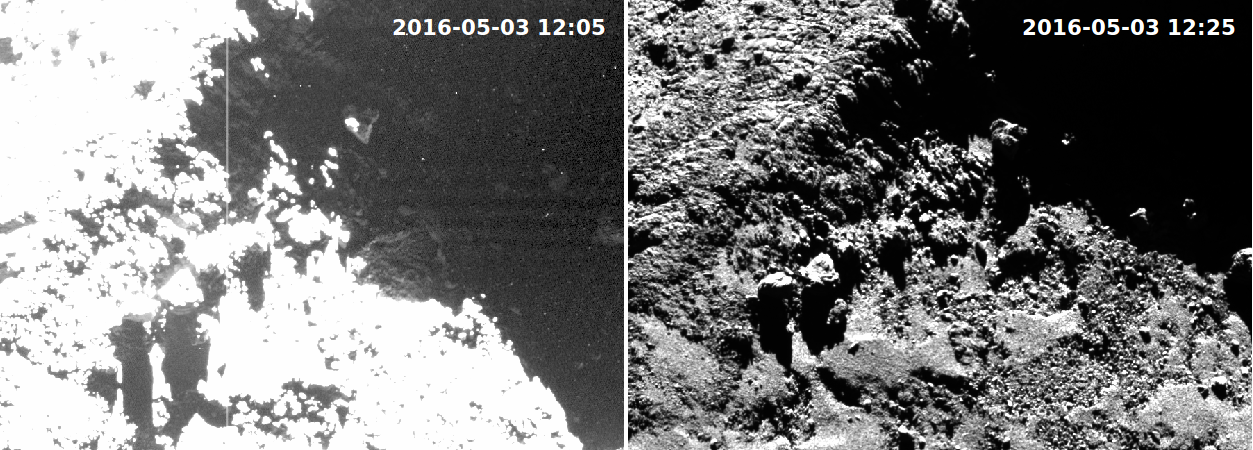}
\caption{Basin F in Imhotep observed with NAC on 2016 January 06 and May 03. The two upper panels show the same image at different brightness scales. The left panel is at a linear scale to show the topography, the right panel at a square-root scale to show the dust plume. The bottom panels show images of May 03 (left: square root scale, right: linear scale) for reference. In the left panel, the circular depression is still in shadow, while on the right it is already exposed to sunlight. Comparison shows that the circular depression is the likely source region of the plume. The observation also shows that a dust plume does not occur at every sunrise.}
\label{fig:jan06}
\end{figure*}

On 2016 January 06, a dust plume similar in shape to that seen on July 3 but a factor 10 less bright was observed near D2. The plume appeared within 10\,minutes from the time when the surface emerged from the shadow cast by the northeastern wall of basin F. Comparison with a similar image obtained on 2016 May 03 shows that the location of the plume is consistent with the southwestern wall of D2 (Fig.~\ref{fig:jan06}). This suggests that the southwestern walls of circular depressions in the Imhotep Basin F may be preferred locations for morning outbursts. However, the plume activity does not occur on every morning, because we saw each depression emerge from shadow without displaying a plume at least twicc (D1: on 2016 May 03, 00:42, and on 2016 June 13, 17:20; D2: on 2016 May 03, 12:25, and on 2016 June 02, 16:56). 

\section{Discussion}
\label{sec:processes}

\subsection{Free sublimation of an icy surface}

\noindent The GIADA data show that a particle of 310\,$\mu$m radius was accelerated to a terminal speed of 1.4\,m\,s$^{-1}$. In the following we examine if this velocity is consistent with the free sublimation of an icy surface.

The acceleration of dust from a small sublimating patch is described in \citep{jewitt_133P}. 
From Table~\ref{tab:temperatures}, we expect equilibrium temperatures between 177 and 189\,K, and corresponding production rates between 1.4$\times$10$^{20}$ and 1.2$\times$10$^{21}$\,s$^{-1}$\,m$^{-2}$. The surface temperature may drop significantly once the optically thick dust plume has formed (ranging from 165 to 180\,K for $A_B$=0.012, 0.6$<\epsilon<$1, and 0$^\circ < \theta <$ 45$^\circ$, and a reduction of the solar irradiation to 50\%).  

\begin{figure}
\includegraphics[width=\columnwidth]{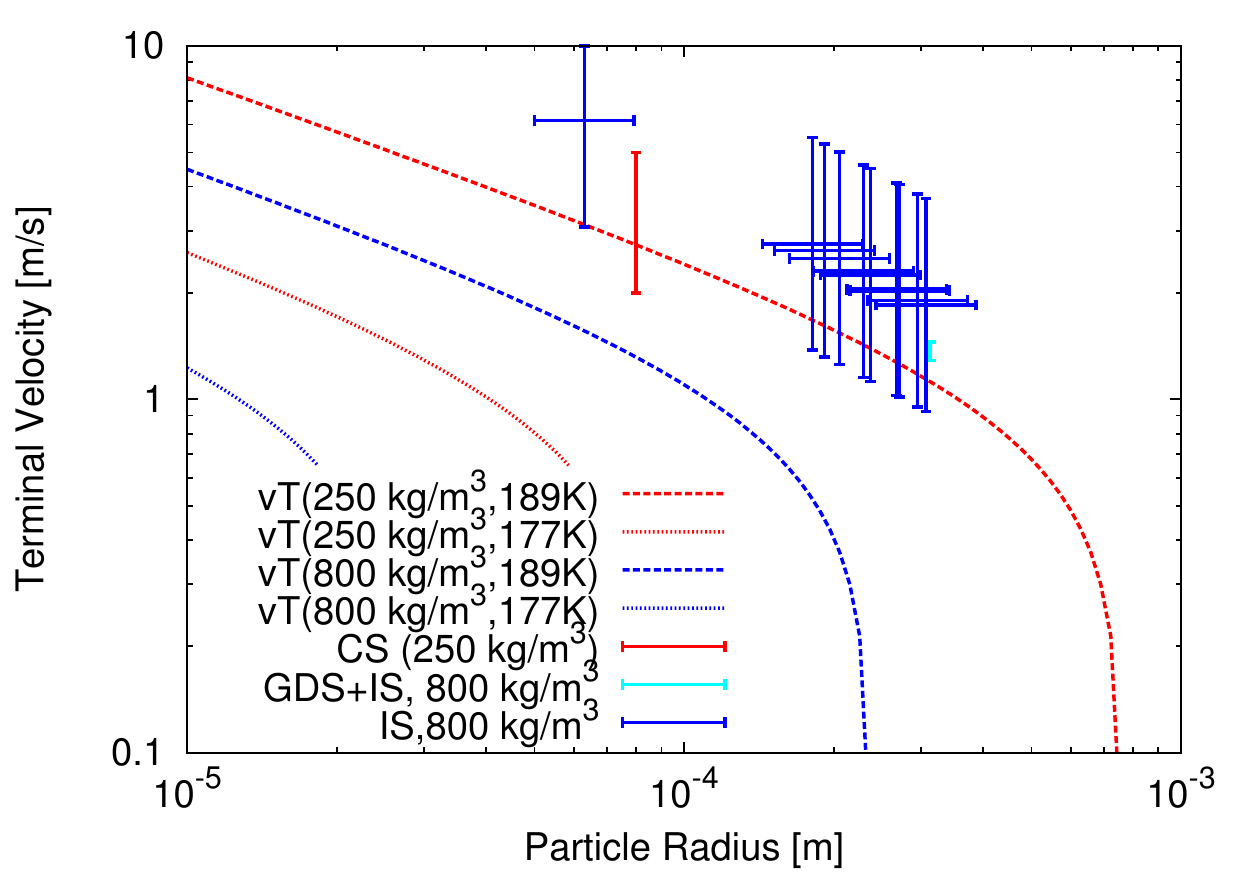}
\caption{Terminal velocity as a function of particle radius derived from GIADA and COSIMA measurements, and expected maximum values $v_T$ for an active patch of 10\,m radius and gas speed of 600\,m\,s$^{-1}$ \citep{jewitt_133P}. The assumed bulk density is colour-coded. Red: 250\,kg\,m$^{-3}$ (COSIMA=CS best estimate), blue: 800\,kg\,m$^{-3}$ (GIADA best estimate). The long dashed lines correspond to the maximum expected surface temperature of 187\,K. The dotted lines correspond to the maximum temperature of 180\,K measured by the MIRO instrument \citep{schloerb-keihm2015}. The particle radii for GIADA data are derived from the measured momentum and the velocity assuming spheres of the indicated bulk density.}
\label{fig:speed_limits}
\end{figure}

Assuming a gas speed of 600\,m\,s$^{-1}$ and an active patch of 10\,m radius (Sec.~\ref{subsec:site}), we plot the size-velocity relation from Eq.~A5 in \citet{jewitt_133P} for bulk densities of 250 and 800\,kg\,m$^{-3}$ and surface temperatures of 177 and 189\,K in Fig.~\ref{fig:speed_limits}. For a density near the lower end and a temperature near the high end of the assumed intervals, the maximum liftable grain size is of order 1\,mm and compatible with the GIADA and STR measurements. The dust velocities are therefore marginally consistent with a sublimating ice patch.

However, the derived dust production rate of (18.4 $\pm$ 10.6)\,kg\,s$^{-1}$ (Sec.~\ref{subsec:mass}) corresponds to (0.06 $\pm$ 0.03)\,kg\,s$^{-1}$\,m$^{-2}$ for a 10\,m-radius patch, a factor(1600 $\pm$ 800) higher than the highest gas production rate listed in Table~\ref{tab:temperatures}. At such a high dust-to-gas mass ratio, the mass loading with dust would significantly influence the gas dynamics and reduce the velocity of both gas and dust. With a dust-to-gas velocity ratio of $\sim$1/200, the dust at terminal velocity would carry only (4 $\pm$ 2)\% of the initial kinetic energy of the gas, but (8 $\pm$ 4)$\times$ its initial momentum, requiring a huge deceleration of the gas. 
It seems therefore highly unlikely that a freely sublimating surface of crystalline water ice can have caused the observed dust plume.

\subsection{Outflow from a pressurised sub-surface reservoir}
An alternative model to explain elevated gas production rates is a pressurised sub-surface reservoir that vents into vacuum through a small opening, such as a crack formed in response to thermal stress. 
%
The tensile strength of the surface layers of comet 67P is estimated to $P_t$ = 3 -- 150\,Pa \citep{vincent-bodewits2015,groussin-jorda2015,basilevsky-krasilnikov2016}, which gives an upper limit to the possible pressure inside the cavity. 

The mass flow rate of gas from a container at pressure $p_{in}$ and temperature $T_{in}$ through a slit of width $a$ and length $l$ into vacuum is given in \citep{sharipov-kozac2009} as
\begin{equation}
\dot{M} = \frac{W \sqrt{m}}{\sqrt{2 \pi k_B}} \frac{p_{in}\,a\,l}{\sqrt{T_{in}}},
\label{eq:mdot} 
\end{equation}
where $m$ is the molecular mass of the gas, $W$ is a dimensionless parameter characteristic of the flow regime (we use $W$ = 1.5 for a viscous flow), and $k_B$ is Boltzmann's constant. 
Substituting $p_{in}$ with the sublimation pressure $P_{subl} (T)$ \citep{fanale-salvail1984} and requiring a minimum total gas production rate of  7.9\,kg\,s$^{-1}$ (to match the minimum dust production rate), we require a temperature  $T_{in}$=260\,K for $l$=20\,m and $a$=1\,m and water vapour, corresponding to an internal pressure of  200\,Pa. This  temperature is significantly higher than the equilibrium temperature derived in Table~\ref{tab:temperatures}. 
For CO$_2$ gas, an internal pressure of  120\,Pa at 180\,K would be sufficient to provide the required gas mass flux, which is easily compatible with a sublimation pressure of $\sim$30,000\,Pa.

Behind the crack, the gas density drops approximately twice as fast as the temperature \citep{sharipov-kozac2009}, while the sublimation pressure drops exponentially with temperature. The gas immediately behind the crack therefore is in a supercooled state and is expected to freeze out, especially in the presence of condensation nuclei such as dust grains. This  could explain the observation of micron-sized water ice particles by Alice.

Assuming that the reservoir is not significantly refilled during the duration of the outburst, that the temperature inside the cavity remains constant, and that the gas inside the cavity can be described by the ideal gas law, the pressure inside the cavity and the mass flux though the crack drop exponentially with an e-folding time, $\tau$, proportional to the volume of the cavity:
\begin{equation}
\tau = \left( \frac{2 \pi m}{k_B T_{in}}\right)^{\frac{1}{2}} \frac{V}{W\,a\,l}.
\end{equation}
For the production rate to drop to 1/1000 of its initial value (maximum liftable grain size of 1\,$\mu$m) in 14 -- 68\,min, we require (120$<\tau<$590)\,s. For water at 250\,K, a volume of {(5 --  24)$\times$10$^{5}$\,m$^3$ is required, corresponding to a sphere of radius (49 -- 84)\,m, or a half sphere of radius (62 -- 105)\,m. For CO$_2$ at 180\,K, the required volume would be a factor 2 smaller. The crack would therefore have extended over a  considerable fraction of the cavity.

We  now estimate how the production rate from the outburst would compare to the overall background level, if CO$_2$ was the driving gas. Assuming a global CO$_2$ production rate of 10$^{25}$\,s$^{-1}$ \citep{fougere-altwegg2016b}, and assuming that this distributes homogeneously over the illuminated hemisphere, we expect a typical CO$_2$ flux of $6 \times 10^{-10}$\,kg\,s$^{-1}$\,m$^{-2}$ at the position of the spacecraft ($\sim$10\,km from the comet centre). For the outburst, we assume that the gas production rate should be comparable to the dust production rate,  18.4\,kg\,s$^{-1}$, and that this gas would distribute homogeneously over a half-sphere, such that at a distance of 8.5\,km, the flux would be  $4 \times 10^{-8}$\,kg\,s$^{-1}$\,m$^{-2}$, a factor  68 above the estimated background. For CO, a similar calculation gives a factor of 110, if the global number production rate was comparable to that of CO$_2$.

If water vapour was the driving species, it must have been  $\sim$80\,K warmer than the surface  equilibrium temperature.
A possible energy source to heat water vapour could be the steady crystallization of amorphous ice in a deeper layer \citep{gonzalez-gutierrez2008}. It is possible that this process supplied a constant rate of sublimation and sufficient heat to build up a pressurised reservoir below a surface layer impenetrable for the vapour \citep{belton-feldman2008}. Such a layer would need to have a thickness of centimeters to decimeters, and could consist of sintered material or of ice frozen out at a depth not reached by the diurnal heat wave. 
The accumulation of a sub-surface gas reservoir due to an internal heat source has been suggested by \citet{belton-feldman2008}. We here suggest a different trigger mechanism (thermal cracking) to explain the coincidence of the outburst with exposure to sunlight. 

An alternative heating process could be the solid state greenhouse effect \citep{matson-brown1989}, where heat is trapped below a surface composed of a visually translucent medium (such as ice or snow) that is opaque in the mid-IR and shows strong forward scattering \citep{hapke1996}. This effect can lead to a significant increase in temperature below the surface, although the magnitude and depth of the effect is strongly model-dependent \citep{davidsson-skorov2002}.

It is also possible that the sealed cavity was heated and filled with vapour around the time of perihelion, when solar irradiation was sufficient \citep{yelle-soderblom2004}, and preserved both temperature and pressure due to a cover layer of low permeability and heat conductivity until it was opened by the crack formation.
Future modelling work will have to investigate if the heating processes outlined above are consistent with the observed properties of the cometary surface and subsurface. 

\subsection{Transition from amorphous to hexagonal ice}
\noindent 
Alternative to a pressurized gas bubble, we propose that amorphous ice may have been present behind the wall R and transformed to crystalline ice when, upon local sunrise on July 3, either a part of the overhanging wall collapsed or a newly formed thermal crack exposed it to solar irradiation. The temperature increase induced by the phase transition can have been sufficient to raise the sublimation rate to a level consistent with the observed dust production rate and velocities.

At temperatures below $T_c$=200\,K and low pressure, ice freezing out from a vapour assumes the metastable crystal structure of cubic ice before transforming to the stable hexagonal ice, and below $T_{as}$=160\,K, amorphous ice can initially form \citep{murphy-koop2005}. Since the face of the wall R was exposed to sunlight only for $\sim$1/6 of a comet rotation and falls into shadow much earlier than the terrain bordering its top edge, it is possible that sublimation continued below the still illuminated surface and the vapour froze out behind the cold face of the wall. 
Since temperatures in the shadow fall easily below 160\,K, this ice could be initially of cubic or even 
amorphous structure, as long as the deposition rate was sufficiently low for the substrate to absorb the considerable latent heat of condensation ($L \sim$51\,000\,J\,mol$^{-1}$ = 2.83$\times$10$^6$\,J\,K$^{-1}$) without a significant rise in temperature.

It is beyond the scope of this paper to explore this possibility in due detail, such that we limit the discussion to simple energetic considerations. Assuming that the subsurface has a warmer ($\sim$100\,K, \citet{desanctis-capaccioni2015}) region below the surface illuminated during local afternoon, and a colder region behind the shortly-illuminated wall, water vapour from the warmer subsurface could diffuse to the colder part and recondense there. The free sublimation rate of crystalline water ice at 100\,K is 1.4$\times$10$^{-17}$ kg\,s$^{-1}$\,m$^{-2}$. This provides a condensation energy flux of 4$\times$10$^{-11}$\,W\,m$^{-2}$. For the local increase in temperature not to exceed $\Delta T$ = 10\,K, and assuming that  at a distance of} $l$=4\,cm behind the condensation front the temperature is not elevated, we obtain a heat flux of $\dot{Q} = k \Delta T \ l = 2.5$\,W\,m$^{-2}$, where k=10$^{-2}$ is the heat conductivity \citep{blum-gundlach2017}. The surrounding material may therefore be able to conduct the latent heat of condensation away from the condensation region without elevating the temperature to a point where amorphous ice could not exist. Even a significantly higher gas flux up to $\dot{Q}/L = 9\times 10^{-7}$\,kg\,s$^{-1}$\,m$^{-2}$ (corresponding to T=170\,K) could be sustained, such that also vapour released in the immediate sub-surface could diffuse to the colder region and recondense there. 

The metastable states transform to hexagonal ice on timescales of minutes to days. The transformation can be accelerated by heating above $T_c$ and $T_{as}$  for cubic and amorphous ice, respectively \citep{murphy-koop2005}. The transformations are exothermic. The latent heat of the transition from cubic to hexagonal ice is (110 $\pm$ 50)\,J\,mol$^{-1}$, and the heat capacity of hexagonal ice at 180\,K is 26\,J\,mol$^{-1}$\,K$^{-1}$ \citep{murphy-koop2005}, such that the transformation would lead to a temperature increase of (4 $\pm$ 2)\,K. The latent heat of the transition from amorphous to crystalline ice is of order 9$\times$10$^{4}$\,J\,kg$^{-1}$ \citep{gonzalez-gutierrez2008} or 1620\,J\,mol$^{-1}$, corresponding to a temperature increase of 62\,K following the phase transition in pure ice. Assuming that the  freshly exposed surface had a radiative equilibrium temperature of $\sim$180\,K (Table~\ref{tab:temperatures}), the temperature would have increased to $\sim$240\,K due to the crystallization. The corresponding sublimation rate would have been 0.033\,kg\,s$^{-1}$\,m$^{-2}$, comparable to the inferred dust production rate of (0.06 $\pm$ 0.03)\,kg\,s$^{-1}$\,m$^{-2}$.

In order to sustain a gas production rate of 0.033\,kg\,s$^{-1}$\,m$^{-2}$ for at least 20\,min, an ice mass of 40\,kg is required. The face of the wall has an area of 20$\times$10\,m$^{2}$ (Fig.~\ref{fig:mar19-jul02}). For a condensation rate of 9$\times$10$^{-7}$\,kg\,s$^{-1}$\,m$^{-2}$, the time required to build the crystallizing area is 3\,days. For a condensation rate of 6$\times$10$^{-9}$\,kg\,s$^{-1}$\,m$^{-2}$, the build-up time would be one year, comparable to the duration of southern summer on 67P.

\section{Summary and conclusion}
\label{sec:summary}
The described observations combine multi-faceted measurements of the outflowing material with detailed information on the surface morphology and composition at the outburst site. Our key findings are:

\begin{itemize}
\item The outburst was located near a northeast-facing wall of $\sim$10m height in the southern hemisphere.
\item The wall emerged from the shadow of a higher, opposite wall after the cometary night 6 minutes before the first detection of the outburst.
\item The surface at the foot of the wall was enriched in water ice before the outburst.
\item A similar dust plume was observed 6 months earlier to originate from a neighbouring depression with similar properties.
\item The dust production was continuous, lasting at least 14 and not longer than 68 minutes.
\item The outburst altered a 10m radius area of the surface and left an icy patch of a projected size of 15$\times$5\,m$^2$.
\item The ejected material comprised sub-micron-sized water ice grains at (12 -- 25)\% of the cross-section, and refractory dust several hundred micron in size.
\item The ejected dust mass was  (6\,500 -- 118\,000)\,kg, corresponding to a layer of  (8 -- 47)\,cm for a 10m-radius patch.
\item The dust production rate was  (18.4 $\pm$ 10.6)\,kg\,s$^{-1}$.
\item For a freely sublimating water ice patch at 3.32\,AU, this would have corresponded to a dust-to-ice mass ratio of  (1600$\pm$800).
\item As such a high mass loading is inconsistent with the observed dust velocities, the free sublimation of water ice alone cannot explain the observed dust production.
\item We conclude that the release of energy stored in the sub-surface must have supported the acceleration of dust.
\end{itemize}

The measurements of July 3 provide reasonably robust evidence that the event was driven by a process more vigorous than the free sublimation of ice, and that some form of energy stored in the sub-surface must have supported direct solar irradiation in accelerating dust. We have discussed two possible forms of such energy storage (a pressurised cavity and near-surface amorphous ice), but the viability of these propositions will have to be tested by future in-depth thermal models and comparison to a larger data set. 

\citet{auger-groussin2015} concluded from the radial pattern of fractures around Basin F, that it formed either by an impact or by the rising of a gas bubble. They interpret nearby roundish features as the walls of ancient gas conduits. It is therefore possible that there still is a gas-filled cavity below Basin F, and that the roundish features inside it are venting tubes still active. 

Primordial amorphous ice has long been suspected to play a significant role in the evolution of the cometary interior and for outbursts \citep[e.g.][]{prialnik-benkhoff2004,prialnik-sarid2008}. We here propose that ice recondensed from a vapour below a badly illuminated surface could initially be amorphous, too, and may cause violent outbursts when eventually exposed to sunlight.

 However, \citet{capria2017} find that primordial amorphous ice can exist as shallow as 1\,m below the cometary surface, such that it is possible that the collapse of a 10\,m high overhanging wall exposed such material.

Near-surface accelerations inconsistent with the free sublimation of water ice have been found in earlier studies \citep{kramer-noack2016,agarwal-ahearn2016}, suggesting that the underlying process may be quite common and significantly contribute to the mass loss in comet 67P. 

\section*{Acknowledgements}
We thank the Rosetta Science Ground Segment at ESAC, the Rosetta Missions Operations Centre at ESOC, and the Rosetta Project at ESTEC for their outstanding work enabling the science return of the Rosetta Mission. Rosetta is an ESA mission with contributions from its Member States and NASA.
The Alice team acknowledges continuing support from NASA's Jet Propulsion Laboratory through contract 1336850. 
COSIMA was built by a consortium led by the Max-Planck-Institut f\"ur Extraterrestrische Physik, Garching, Germany in collaboration with Laboratoire de Physique et Chimie de l'Environnement et de l'Espace, Orl\'eans, France, Institut d'Astrophysique Spatiale, CNRS/Universit\'e Paris Sud, Orsay, France, Finnish Meteorological Institute, Helsinki, Finland, Universit\"at Wuppertal, Wuppertal, Germany, von Hoerner und Sulger GmbH, Schwetzingen, Germany, Universit\"at der Bundeswehr, Neubiberg, Germany, Institut f\"ur Physik, Forschungszentrum Seibersdorf, Seibersdorf, Austria, Space Research Institute, Austrian Academy of Sciences, Graz, Austria and is led by the Max-Planck-Institut f\"ur Sonnensystemforschung, G\"ottingen, Germany. The support of the national funding agencies of Germany (DLR, grant 50 QP 1302), France (CNES), Austria, Finland and the ESA Technical Directorate is gratefully acknowledged.
GIADA was built by a consortium from Italy and Spain under the scientific responsibility of the Universit{\`a} di Napoli ``Parthenope'' and INAF-OAC. GIADA was operated by the Istituto di Astrofisica e Planetologia Spaziali of INAF. GIADA development for ESA has been managed and funded by the Italian Space Agency (ASI-INAF agreements I/032/05/0 and I/024/12/0) with a financial contribution by MEC/ES.
OSIRIS was built by a consortium of the Max-Planck-Institut f{\"u}r Sonnensystemforschung, G{\"o}ttingen, Germany, CISAS University of Padova, Italy, the Laboratoire d'Astrophysique de Marseille, France, the Instituto de Astrof\`{\i}sica de Andalucia, CSIC, Granada, Spain, the Research and Scientific Support Department of the European Space Agency, Noordwijk, The Netherlands, the Instituto Nacional de T\`ecnica Aeroespacial, Madrid, Spain, the Universidad Polit{\`e}chnica
de Madrid, Spain, the Department of Physics and Astronomy of Uppsala University, Sweden, and the Institut f{\"u}r Datentechnik und Kommunikationsnetze der Technischen Universit{\"a}t Braunschweig, Germany. The support of the national funding agencies of Germany (DLR), France(CNES), Italy(ASI), Spain(MEC), Sweden(SNSB), and the ESA Technical Directorate is gratefully acknowledged.
We acknowledge the contributions made by Alan Stern and Eric Schindhelm to the Alice measurements, and by Hans Rickman to the OSIRIS measurements.
We thank the referee, Gal Sarid, for his comments that helped to improve the manuscript.




\bibliographystyle{mnras}
\bibliography{refs} 




\appendix

\section{Additional OSIRIS images}

\begin{figure*}
\includegraphics[width=\textwidth]{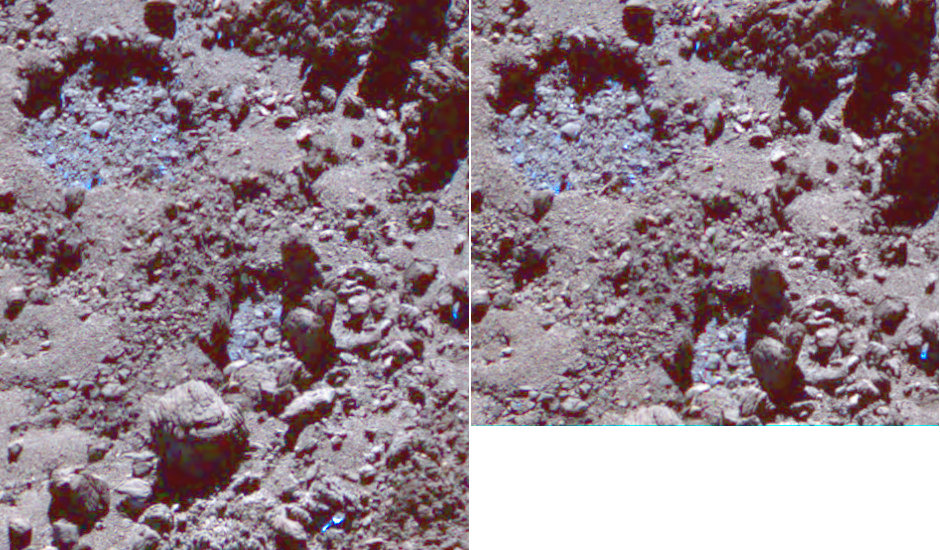}
\caption{False colour rgb images composites of the outburst site on 2016 July 02, $\sim$10\,h before the outburst, composed from three OSIRIS/NAC images in NIR (red channel), orange (green channel) and blue (blue channel) filters. The left panel (identical to the right panel in Fig.~\ref{fig:mar19-jul02}) shows observations obtained at 21:26, the right panel is based on observations from 21:36.} 
\label{fig:july2}
\end{figure*}

\begin{figure*}
\includegraphics[width=\textwidth]{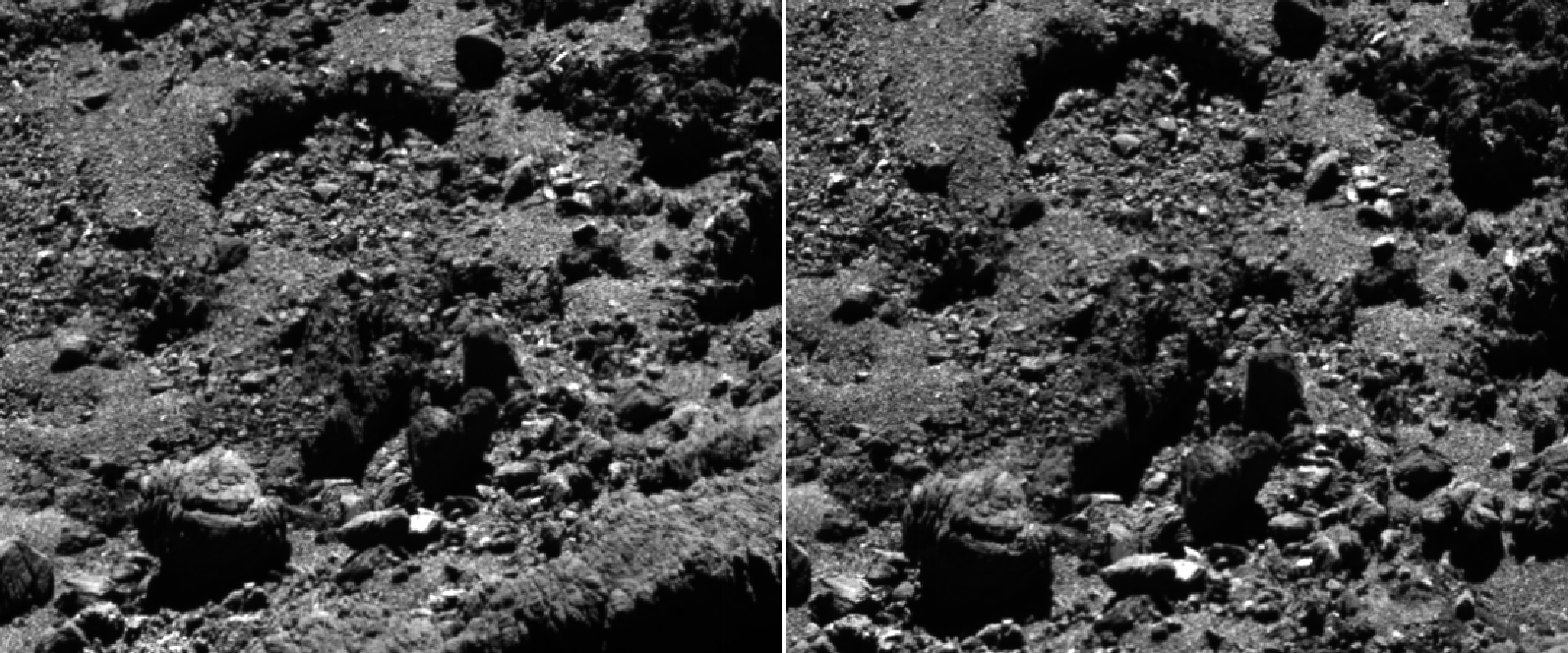}
\includegraphics[width=\textwidth]{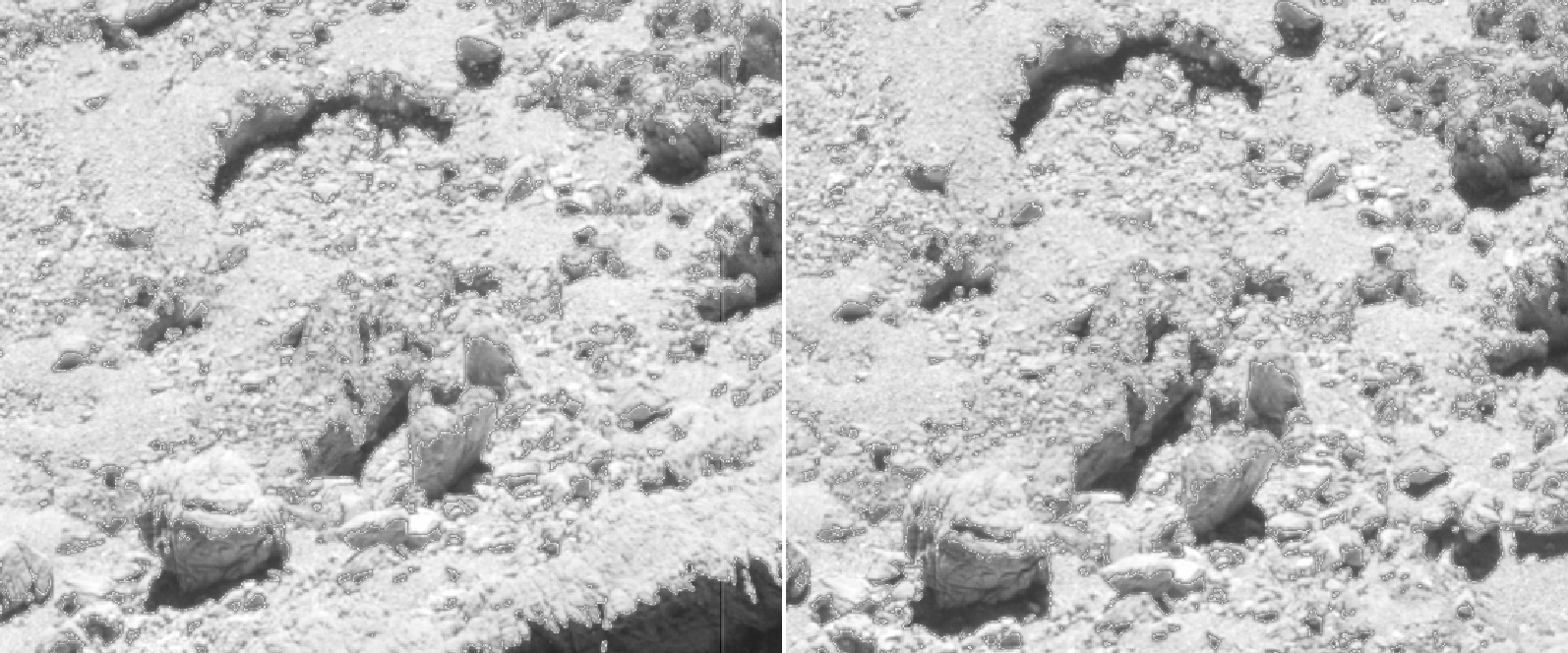}
\caption{The outburst site D1 and the neighbouring depression D2 on 2016 March 19 at UT 21:26 (left) and 21:46 (right). The upper panels show the image at a linear scale ranging from 0 (black) to 10$^{-4}$ W\,m$^{-2}$\,nm$^{-1}$\,sr$^{-1}$. The lower panels show the same images at a logarithmic scale between 0 and 10$^{-3}$ W\,m$^{-2}$\,nm$^{-1}$\,sr$^{-1}$ for illuminated regions and at linear scaling between 0 and 4$\times$10$^{-5}$ W\,m$^{-2}$\,nm$^{-1}$\,sr$^{-1}$ for the shadowed regions to maximise visibility of the indirect illumination from the sunlit surface. These are to our knowledge the best images of the northeastern face of the wall R and its counterpart in D2. The projected size of the shadowed face of R is 17$\times$7\,m. These are lower limits due to the unknown angle of projection. The bottom left panel is identical to the left panel of Fig.~\ref{fig:mar19-jul02}.}
\label{fig:march19}
\end{figure*}

\begin{figure*}
\includegraphics[width=\textwidth]{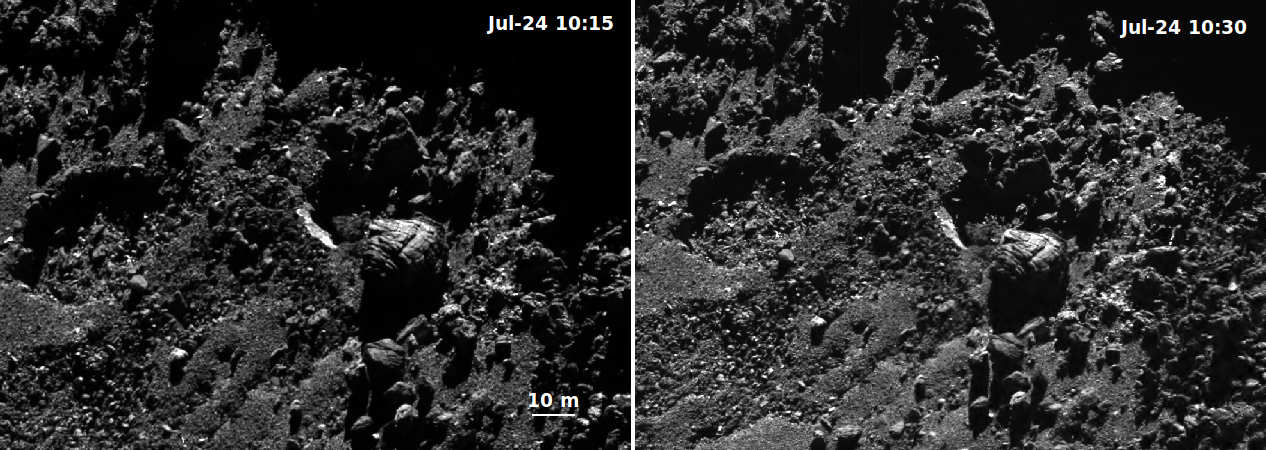}\\[0.05\baselineskip]
\includegraphics[width=\textwidth]{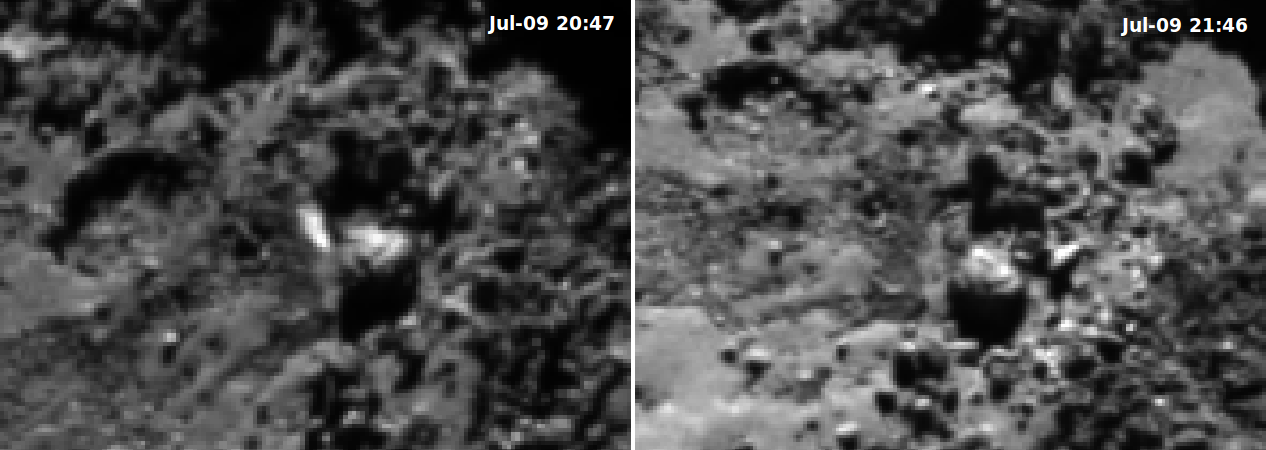}\\[0.05\baselineskip]
\includegraphics[width=\textwidth]{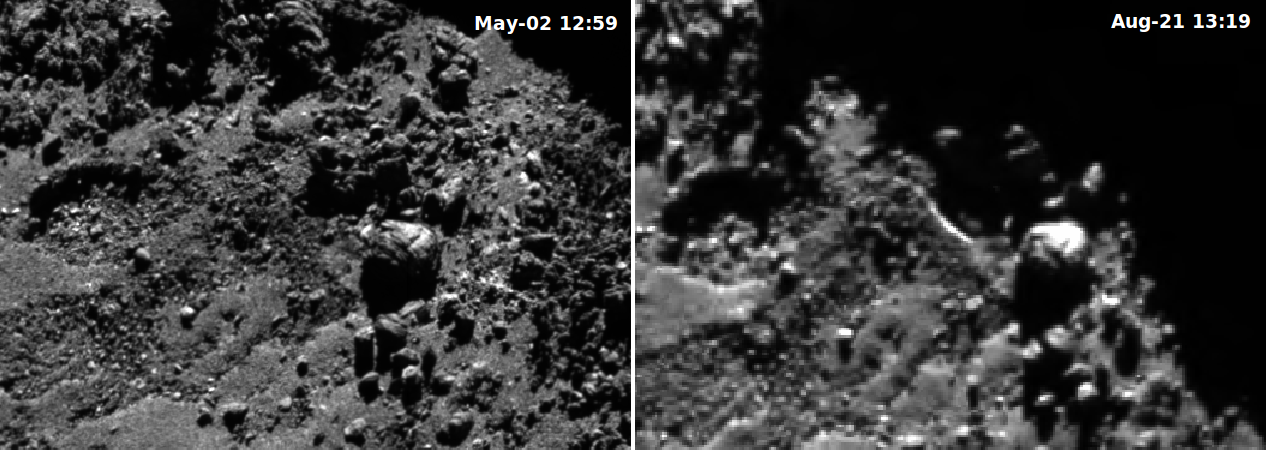}
\caption{Images of the outburst site, obtained between 2016 May 02 and August 21. The observation parameters are listed in Table~\ref{tab:image_parameters}. A bright patch at the site of the outburst was detected on July 09, 24, and August 21. The patch quickly enters shadow between 11:34\,h and 13:23\,local time, indicating a steep wall. The image of May 02 does not show the icy patch. It is the pre-outburst image matching most closely the viewing and illumination conditions of those showing the icy patch, although it was taken from a more shallow (by 3$^\circ$) position. We use this image for a close comparison between pre- and post-outburst state of the source region in Fig.~\ref{fig:july24_details}}
\label{fig:ice_patch}
\end{figure*}


\bsp	
\label{lastpage}
\end{document}